%% file: mana050801v4-prob_of_prob.tex
\RequirePackage{txfonts}\normalfont     

\documentclass[11pt,
onecolumn,oneside,a4paper,article
,italian,frenchb,german,
swedish,british%
]{memoir}

\newcommand{\usetxfonts}{\usepackage{txfonts}}
\input{preamblemem.tex}

\alleugreek

\usepackage{paralist}

\usepackage[hypertex,breaklinks,pdfauthor={P G L  Porta Mana}]{hyperref}

\usepackage[numbers,sort&compress]{natbib}
\usepackage{memhfixc}

\setlxvchars
\setxlvchars

\reparticle
\checkandfixthelayout
\fixpdflayout

\copypagestyle{manaart}{plain}
\makeheadrule{manaart}{\headwidth}{\normalrulethickness}
\makeoddhead{manaart}{\small\textsc{Porta Mana, M\aa{}nsson,
  Bj\"ork}}{}{\small\textit{`Plausibilities of plausibilities': the
  circumstance approach}}

\newenvironment{acknowledgements}{\chapter*{Acknowledgements}\addcontentsline{toc}{chapter}{Acknowledgements}}{\par}

\newenvironment{rem}{\begin{remark}}{\qee\end{remark}}

\theoremstyle{plain}

\DeclareMathOperator{\N}{N}
\DeclareMathOperator{\Br}{Br}
\DeclareMathOperator{\truth}{T}
\DeclareMathOperator{\convh}{conv}

\newcommand{\persa}{Cecily}
\newcommand{\persb}{Gwendolen}
\newcommand{\persc}{Jack}
\newcommand{\persd}{Algernon}

\newcommand{\znorm}[1]{\lVert#1\rVert_{\infty}}

\newcommand{\zmu}{\mu}
\newcommand{\zsi}{\sigma}

\newcommand{\zM}{M}
\newcommand{\zR}{R}
\newcommand{\zr}{r}

\newcommand{\zn}{n}
\newcommand{\zm}{m}

\newcommand{\zS}{S}
\newcommand{\zSp}{C}
\newcommand{\zSpa}{\zSp_\text{C}}
\newcommand{\zSpb}{\zSp_\text{G}}
\newcommand{\zSpc}{\zSp_\text{J}}
\newcommand{\zSpd}{\zSp_\text{A}}

\newcommand{\zx}{x}

\newcommand{\zq}{q}
\newcommand{\zqh}{\zq_\text{h}}

\newcommand{\zqq}{\bm{\zq}}
\newcommand{\zqqq}{\bar{\zqq}}
\newcommand{\zqql}{{\zqq^\text{L}}}
\newcommand{\zqqn}{{\zqq^\text{N}}}

\newcommand{\zrA}{\varDelta}
\newcommand{\zQE}{{\varGamma_{N}}}
\newcommand{\zeqcl}{{\zqqq}\sptilde}
\newcommand{\zQQ}{\varGamma}
\newcommand{\zQQQ}{\overline{\varGamma}}
\newcommand{\zQQE}{{\varGamma_{\infty}}}

\newcommand{\zrB}{\omega}

\newcommand{\zf}{f}

\newcommand{\zE}{E}

\newcommand{\zMl}{\zM^\text{L}}
\newcommand{\zMn}{\zM^\text{N}}

\newcommand{\zRnu}{\zR^\text{N}}
\newcommand{\zRa}{\zR^\text{L}_\text{a}}
\newcommand{\zRb}{\zR^\text{L}_\text{b}}
\newcommand{\zRo}{\zR^\text{N}_\text{1}}
\newcommand{\zRt}{\zR^\text{N}_\text{2}}
\newcommand{\zRH}{\zR_\text{h}}
\newcommand{\zRT}{\zR_\text{t}}

\newcommand{\zSpexa}{\zSp_{(2\text{a1},0\text{a2},1\text{b1},1\text{b2})}}
\newcommand{\zSpexb}{\zSp_{(1\text{a1},1\text{a2},2\text{b1},0\text{b2})}}

\newcommand{\zSexab}{\zS_{((1/2,1/2),\, (3/4,1/4))}}
\newcommand{\zSql}{\zS_{(\zqql,\zqqn)}}
\newcommand{\sumqq}{\sum_{\zqqq}}
\newcommand{\sumqql}{\sum_{\zqqq'}}
\newcommand{\ka}{k}
\newcommand{\kb}{l}

\newcommand{\zSq}{\zS_{\zqq}}
\newcommand{\zSqp}{\zS_{\zqq'}}
\newcommand{\zSqq}{\zS_{\zqqq}}
\newcommand{\zSqql}{\zS_{\zqqq'}}

\newcommand{\zih}{{\text{h}}}
\newcommand{\zit}{{\text{t}}}

\newcommand{\zp}{p_{\zS}}
\newcommand{\zpp}{p_{\zS'}}
\newcommand{\zppp}{p_{\zS''}}

\newcommand{\zJ}{J}

\newcommand{\zIc}{I_\text{co}}
\newcommand{\zIco}{J_\text{co}}

\newcommand{\zIn}{\zI_N}
\newcommand{\zI}{I}

\newcommand{\zIi}{I_{\infty}}

\pretitle{\begin{center}\LARGE\bfseries}
\predate{\begin{center}(Dated: }
\postdate{)\end{center}}

\providecommand{\affiliation}[1]{\textit{\small#1}}
\providecommand{\pacs}[1]{{\small\textsc{PACS} numbers: #1}}
\providecommand{\msc}[1]{{\small\textsc{MSC} numbers: #1}}
\providecommand{\email}[1]{\texttt{\href{mailto:#1}{#1}}}

\title{`Plausibilities of plausibilities':\\ an approach
  through circumstances
\\[3\jot]
\textnormal{\large Being part I of\\``From `plausibilities of
    plausibilities' to state-assignment methods''}}

\author{\firstname{P. G. L.} \surname{Porta Mana},\thanks{Email: \email{mana@kth.se}}
\quad
\firstname{A.} \surname{M{\aa}nsson},
\quad
\firstname{G.} \surname{Bj\"{o}rk}
\\
\affiliation{Kungliga Tekniska H\"ogskolan, Isafjordsgatan
  22, SE-164\,40 Stockholm, Sweden}
}

\date{29 April 2007}

\begin{document}
\bibliographystyle{apsrevmananum} 

\setlength{\droptitle}{-3\onelineskip}

\setcounter{chapter}{-1}

\maketitle

\begin{abstract}
  Probability-like parameters appearing in some statistical models, and
  their prior distributions, are reinterpreted through the notion of
  `circumstance', a term which stands for any piece of knowledge that is
  useful in assigning a probability and that satisfies some additional
  logical properties. The idea, which can be traced to Laplace and Jaynes,
  is that the usual inferential reasonings about the probability-like
  parameters of a statistical model can be conceived as reasonings about
  equivalence classes of `circumstances' --- \viz, real or hypothetical
  pieces of knowledge, like \eg\ physical hypotheses, that are useful in
  assigning a probability and satisfy some additional logical properties
  --- that are \emph{uniquely indexed} by the probability distributions
  they lead to. 
  \\[2\jot]
  \pacs{02.50.Cw,02.50.Tt,01.70.+w}\\
  \msc{03B48,62F15,60A05}
\end{abstract}


\epigraph{\emph{If you can't join 'em,\\
        join 'em together.}}%
{}

\chapter{Introduction}
\label{sec:intro}

In the present first study we offer an alternative point of view on, or a
re-interpretation of, probability-like parameters and `probabilities of
probabilities', two objects that appear in connexion with statistical
models. This also provides a re-interpretation of some kinds of inverse
methods, for which we develop a simple and general logical framework. This
point of view, which we think is basically Laplace's but uses an idea
presented \emph{in nuce} in some work by Jaynes, is alternative to both
that based on the distinction between `physical' and `subjective'
probabilities, and that based on de~Finetti's theorem. This point of view
and the ensuing inverse-method framework have applications in physical
theories, as will be shown in following
studies~\citep{portamana2007b,portamana2007}.


\medskip

Our study and results are based exclusively on probability
theory; we do not use entropy notions, for example. For
us, `probability' means simply \emph{plausibility}, and
the following conceptual proportion holds:\footnote{Does
  also Wittgenstein mean something of the kind when he
  writes `Probability theory is only concerned with the
  state of expectation in the sense in which logic is with
  thinking'~\citep[\sect~237]{wittgenstein1929_t1998}\citep[p.~231]{wittgenstein1933b_t1993}?
See also Johnson~\citep[pp.~2--3]{johnson1932}.}
\begin{multline*}
  \label{eq:concept_eq}
  {\text{Plausibility calculus}} : {\text{Everyday notion
      of `plausibility'}}
  ={}\\
  {\text{Logical calculus}} : {\text{Everyday notion of
      `truth'}}.
\end{multline*}
We thus take 
the licence to adopt the term \emph{plausibility}
henceforth\footnote{Also used by
  Kordig~\citep{kordig1978}.} --- with no need to define
what it means, any more than it is usually done with
`truth'.\footnote{Is truth objective or subjective?  Can
  truth be `operationally' defined? Can the truth of a
  proposition be tested? --- `Of course! to test the truth
  of \prop{This hat is brown} I only need to look at the
  hat!'. Well, provided \eg\ that you are not dreaming or
  having hallucinations; and how do you test \emph{that}?
  Going backwards, in the end you arrive at some
  proposition which you simply \emph{assume} ---
  unconsciously, by convention, by agreement, by caprice
  --- and cannot `test'. `Subjectivity' lurks no less in
  logic than in plausibility theory --- and is no less
  \emph{un}interesting in plausibility theory than it is
  in logic.} (Our study can in any case be easily
`translated' into degrees-of-belief, credence, or
similar terms.)

The notation $\pr(A \cond \zI)$
\citep{johnson1932,johnson1932b,johnson1932c,koopman1940,koopman1940b,koopman1941,cox1946,cox1961,adams1975,cox1979,jeffreys1931_r1957,jeffreys1939_r1998,jaynes1954,jaynes1959,jaynes1994_r2003,hailperin1996,adams1998,gregory2005}
(\cf\
also~\citep{carnap1950_r1962,gaifman1964,scottetal1966,hacking1967,gaifman1979,gaifmanetal1982,maher1999,gaifman2003,hajek2003,
  maher2004,fitelsonetal2005,fitelson2005,fitelsonetal2005b,maher2006})
will denote the plausibility of the statement $A$ in the \emph{context}
described by the proposition $\zI$.\footnote{The context could also be
  called `condition' or `situation'; Johnson calls it
  `supposal'~\citep{johnson1932} ; Jeffreys simply
  `data'~\citep{jeffreys1939_r1998}. In the notation above, there is a
  relation (in the sense in which there is a relation between `certain' and
  `true') between the expressions `$\pr(A \cond \zI)=1$' and `$\zI \models
  A$' (especially when this is used as in situational logic; see
  \eg~\citep{barwise1985,barwise1989,restall1996,gaifman2002}). The latter
  could also be suggestively written `$\truth(A\cond \zI)=1$'. The
  differences between the formalisms of logic and plausibility theory lie
  essentially in the fact that our everyday use of truth can be effectively
  (but not exclusively) \emph{modelled} through a dichotomic set like
  $\set{0,1}$ (or $\set{\top, \bot}$ or $\set{\text{T}, \text{F}}$),
  whereas our everyday use of plausibility can be effectively
  \emph{modelled} through an ordered continuum like $\clcl{0,1}$ (or
  $\clcl{0,+\infty}$ or $\clcl{-\infty,+\infty}$, see \eg\
  Tribus~\citep[pp.~26--29]{tribus1969},
  Jaynes~\citep[\chap~2]{jaynes1994_r2003}, Cox~\citep{cox1946}). This has
  important implications, like the fact that a purely syntactic approach to
  plausibility, in the guise of the logical calculus, is near to
  worthless~\citep{gaifman_pc2006}. The parallel between plausibility
  theory and truth-functional logic suggests also another point. We do not
  require of logic, when put to practical use, that it should also provide
  us with the initial truth-value assignments (the `assumptions'). Why
  should we have an analogous requirement on plausibility theory with
  regard to initial plausibility-value assignments instead?} We shall also
say that $\zI$ `leads to' or `yields' a given plausibility of $A$, but no
particular meaning is intended with these two verbs. Associated
plausibility densities will be denoted by $x \mapsto p_A(x \cond \zI)$; the
term `distribution' will be used for `density' sometimes. Other symbols and
notations are used in accordance with \textsc{ISO}~\citep{iso1993} and
\textsc{ANSI/IEEE}~\citep{ieee1993} standards.

In another study~\citep{portamanaetal2007} we analyse the question of
assigning plausibilities to unknown `events' (\eg, measurement outcomes)
from knowledge of `similar events'; a problem which is connected to
induction. The key point is the formalisation, within probability theory,
of the notion of `similar event'. This we do through the framework and the
interpretation presented in the first note. We do not use the idea of
exchangeability --- and \emph{infinite} exchangeability in particular ---
which is used in Bayesian theory for the same purpose; but there are known
strong connexions and analogies with its mathematics and some of its
results. In fact, we try to persuade the reader that our approach touches
the core idea from which exchangeability also springs.

In a third note~\citep{portamana2007b,portamana2007} it will be shown that
the inferential point of view presented here and
in~\citep{portamanaetal2007} finds applications in physical theories, like
classical and quantum mechanics, an example being state
reconstruction~\citep{maanssonetal2006,maanssonetal2007}. The framework
presented subsumes and re-interprets known techniques of quantum-state
assignment (or `retrodiction' or `reconstruction') and tomography, and
offers alternative approaches to analogous techniques in classical
mechanics.



\chapter{Statistical models and `probabilities of probabilities'}
\label{sec:prob_of_prob}

A statistical model is, roughly speaking, a plausibility distribution whose
numerical values depend on parameters (for a critical discussion of more
rigorous or useful definitions see~\citep{mccullagh2002,besagetal2002}). An
example is the ubiquitous normal distribution
\begin{equation*}
x \mapsto \N(x \cond \zmu, 
1/\zsi^2)
\defd
\frac{1}{\sqrt{2\piup}\, \zsi}
\exp\biggl[-\frac{(x-\zmu)^2}{2\zsi^2}\biggr]
\end{equation*}
whose parameters are the expectation $\zmu$ and the
variance $\zsi$. Another example, one in which we shall be
especially interested in this paper, is the `generalised
Bernoulli' model $i \mapsto \Br(i \cond \zqq)$, which
gives the plausibility distribution for a set of $\zm$
mutually exclusive and exhaustive propositions
$\set{\zR_1, \dotsc, \zR_\zm}$, hereafter called
\emph{outcomes}, 
\begin{equation}
  p(\zR_i \cond \zqq) = \Br(i \cond \zqq)
  \defd \zq_i,\label{eq:bernoulli_general}
\end{equation}
depending on a set of parameters $\zqq\defd (\zq_1,
\dotsc, \zq_\zm)$ which belong to a simplex of appropriate
dimensionality:
\begin{equation}
  \zqq\in\zrA \defd
  \set{(\zx_i) \st \zx_i\ge 0,
    {\textstyle\sum_i}\zx_i=1}.\label{eq:simplex_def}
\end{equation}
(This model is apparently called `discrete model'.
Since this name is too anonymous and the model
reduces to the Bernoulli one for $\zm=2$, we opted for 
`generalised Bernoulli' instead.)

The parameters of a statistical model are sometimes
regarded as `unknown', and a plausibility distribution
(more precisely, a density) for them is therefore
introduced. This distribution, usually called `prior
distribution' or simply `prior', is used in calculations
for a variety of purposes; two in particular interest us
here and in the following papers. (1) A parameter-free
plausibility distribution for the outcomes can be obtained
integrating the product of the prior and the statistical
model in respect of the parameters, \ie, by
marginalisation. In the case of the Bernoulli model, \eg,
introducing a prior $\zqq\mapsto \zf(\zqq)$ ($\zf$~being
of course a normalised positive generalised function) one
obtains the parameter-free distribution
\begin{equation}\label{eq:marginbernoulli}
  p(\zR_i) = \int_\zrA \Br(i \cond \zqq)\, \zf(\zqq)\, \di\zqq
  \equiv
  \int_\zrA \zq_i\, \zf(\zqq)\, \di\zqq.
\end{equation}
(2) In so called `inverse methods' or `problems' (\cf\
Dale~\citep[\sects~1.2,~1.3]{dale1991_r1999}), the prior
is used in the formula of Bayes' theorem to obtain an
`updated' plausibility distribution for the parameters,
conditional on knowledge of some outcome. The resulting
distribution is called `posterior (distribution)'. In the
case of the Bernoulli model, \eg, from the prior
$\zqq\mapsto \zf(\zqq)$ and knowledge of
the outcome $\zR_i$ one obtains the posterior
\begin{equation}\label{eq:inverse_bernoulli}
  \zqq\mapsto
\zf(\zqq \cond \zR_i) =
  \frac{p(\zR_i \cond \zqq)\, f(\zqq)}
  {\int_\zrA p(\zR_i \cond \zqq')\, f(\zqq')\,\di\zqq'}.
\end{equation}

\medskip

Such practices are at least as old as
Bayes~\citep{bayes1763,laplace1774,laplace1812_r1820}. Related and
unrelated historical information can be found \eg\ in Dale's
book~\citep{dale1991_r1999} and some nice essays by
Hacking~\citep{hacking1971,hacking1971b,hacking1971c}; see also Jaynes'
discussion~\citep[\chap~18]{jaynes1994_r2003}. Old, though apparently not
as old as Bayes, 
is also the question: how to interpret statistical-model parameters like
$\zqq$ and their prior distributions? The problem is that the parameters
$(\zq_i)$ look like plausibilities, since their values are identical to the
plausibilities of the outcomes $\set{\zR_i}$ as
\eqn~\eqref{eq:bernoulli_general} shows, and that the prior $f$ looks
therefore like a `plausibility (distribution) of a plausibility' --- a
redundant notion. This question, combined with the related issues on the
interpretation of `probability', has led to many philosophical debates; see
\eg, amongst the vast literature on
this,~\citep{marschaketal1975,moslehetal1996}. The importance of the
interpretative question is not merely philosophical, however. Different
interpretations can lead to different conceptual and mathematical
approaches --- and thus to different solutions --- in the investigation of
concrete problems. This is particularly true for elaborate statistical
models, like those connected to physical theories.

Two main interpretations appear to be in vogue. Many
statisticians, logicians, and physicists, on the one hand,
speak about `subjective' and `physical' probabilities (or
`propensities'~\citep{gillies2000}). For them the notion
of a `probability of a probability' poses no problems,
since it means something like `the subjective probability
of a propensity'.
The very idea of `estimating a probability' implies such
kinds of interpretation; \cf~\eg\ Good~\citep{good1965},
especially the title and chapter~2,
Jamison~\citep{jamison1970}, or
Tintner~\citep{tintner1941}.

For pious Bayesian or `de~Finettian' devotees, on the
other hand, which conceive probability as `degree of
belief', the notion of a `degree of belief in a degree
of belief' is redundant or even meaningless. The Bayesian
are notoriously rescued from philosophical headaches by
de~Finetti's celebrated theorem and other similar
ones~\citep{definetti1937,definetti1938,hewittetal1955,heathetal1976,lindleyetal1976,diaconis1977,georgii1979,link1980,diaconisetal1980,jaynes1986c,diaconisetal1987,ladetal1990,bernardoetal1994,caves2000,caves2000c},
by which parameters like $\zqq$ and functions like $\zqq
\mapsto \zf(\zqq)$ are introduced as mere
\emph{mathematical devices} --- \ie, not plausibilities or
degrees of belief!\ --- that need not be directly
interpreted. See Bernardo and
Smith~\citep[\chap~4]{bernardoetal1994} for a neat
presentation of this point of view.  Interpretative issues
like the Bayesian's are also shared by those who thinks in
terms of `logical probabilities'~\citep{keynes1921} or,
like we, simply in terms of `plausibilities'.

Here we present, discuss, and formalise still another interpretation ---
let us call it the `circumstance interpretation' for definiteness' sake,
for reasons that will be apparent in the next section ---
in which functions like $\zqq\mapsto \zf(\zqq)$ do represent plausibility
distributions, \ie, they are not mere mathematical devices, but the notion
of `plausibility of a plausibility' is nevertheless completely avoided.
This interpretation combines two ideas by which Jaynes tried to make sense
of `plausibility-like' parameters: one is very briefly formulated
in~\citep[p.~11]{jaynes1986d_r1996}, and can possibly be read also in
de~Finetti~\citep[\sect~20]{definetti1931}; the other --- the idea of an
`$A_p$ distribution' --- appears in the various versions of his book on
probability
theory~\citep[lect.~18]{jaynes1954}\citep[lect.~5]{jaynes1959}\citep[\chap~18]{jaynes1994_r2003}.
A similar interpretation is also proposed and discussed by Mosleh and
Bier~\citep{moslehetal1996}. Caves also discusses, and criticises, a
similar idea~\citep{caves2000c,cavesetal2002}. It really seems to us,
however, that this interpretation is basically what Laplace had in
mind~\citep{laplace1774}, if we read his `\french{causes}' more generally
as `circumstances'.

Instead of trying to summarise this interpretation in
abstract general terms that would very likely only appear
obscure at this point, we prefer to invite the reader to proceed
to the simple and concrete example of the next section,
just a coin toss away.  The example will allow us to
introduce the basic idea, along with some
terminology. Then another, more elaborate example
(\sect~\ref{sec:second_example}) follows, to further
expand the main idea. This is then abstracted and
generalised (\sect~\ref{sec:gener_rem}). 
Some important
remarks are scattered throughout this note.

\chapter{Interpreting plausibility-like parameters as
  `indexed circumstances': introductory example}
\label{sec:introexample}

\section{Context and circumstances}
\label{sec:contex_circum}


A coin has been tossed, the outcome unknown to us. We want
to assign plausibilities to the outcomes `head', $\zRH$, and `tail', $\zRT$. The old
recipe says to compute ``\french{le rapport du nombre
  des cas favorables \`a celui de tous les cas
  possibles}''~\citep{laplace1812_r1820}. This is seldom
of much help: Which are \emph{the} cases?\ at which depth
should the situation be analysed?  And what if these cases
are not equally plausible?

But why not analyse the situation in terms of some set of
`cases' anyway? \emph{Some} set, not \emph{the} set. And
their plausibilities can be assigned by some other
means. We do not want the ultimate analysis, just
\emph{an} analysis.

In our case, suppose that the knowledge of the situation,
which constitutes the context $\zIc$, says that either
\persa\ or \persb\ or \persc\ or \persd\ tossed the
coin. Let us call these the four possible
\emph{circumstances} of the coin toss and denote them by
$\zSpa$, $\zSpb$, $\zSpc$, $\zSpd$. The context could thus
be analysed as the conjunction $\zIc = \zIco \land (\zSpa
\lor \zSpb \lor \zSpc \lor \zSpd)$, for some
`sub-context' $\zIco$.

Each circumstance says also something more about the
respective person, which helps us in assigning the
conditional plausibilities:\footnote{\Cf\ Laplace's
  \french{Probl\`eme~II}~\citep{laplace1774}.}
\begin{itemize}
\item[$\zSpa$:] \persa\ is a magician and skilled coin-tosser that
  always like to produce the outcome `head'. If we knew
  that she had tossed the coin, we would
  assign the distribution of plausibility
\begin{equation}
  \label{eq:persa}
 \Bigl( \pr(\zRH \cond \zSpa \land \zIc),\;
  \pr(\zRT \cond \zSpa \land \zIc)\Bigr) = (1,0)
\end{equation}
for the outcomes.
\item[$\zSpb$:] \persb, on the other hand, has no such particular
  skills, so if it were her who had tossed the coin we
  would assign the plausibility distribution
  \begin{equation}
  \label{eq:persb}
 \Bigl(\pr(\zR_i \cond \zSpb \land \zIc) \Bigr) =
\Bigl(\tfrac{1}{2},\tfrac{1}{2} \Bigr),
\end{equation}
with $i =\text{h}, \text{t}$ here and in the
following.
\item[$\zSpc$:] On \persc\ we know nothing whatsoever. He could be
  skilled or unskilled in coin-tossing, a trickster or an
  absolutely earnest person. If we knew he had tossed the
  coin we could but assign the distribution
\begin{equation}
  \label{eq:persc}
  \Bigl(\pr(\zR_i \cond \zSpc \land \zIc)
\Bigr) =
  \Bigl(\tfrac{1}{2},\tfrac{1}{2} \Bigr).
\end{equation}
\item[$\zSpd$:] Finally, we know that \persd\ had been carrying a
  double-headed coin, which he would exchange with the
  original one if asked to toss it. So we assign the
  plausibilities
  \begin{equation}
  \label{eq:persd}
 \Bigl( \pr(\zR_i \cond \zSpd \land \zIc)\Bigr) = (1,0)
\end{equation}
in case he had made the coin toss.
\end{itemize}

\begin{rem} \label{rem:circum_no_causes} It is clear that
  not all the circumstances above express
  `causes'~\citep{laplace1774} or
  `mechanisms'~\citep[lects.~16,~17]{jaynes1954}\citep[lect.~5]{jaynes1959}\citep[\chap~18]{jaynes1994_r2003}\citep{caves2000c}
  which `determine' the respective plausibility
  distributions. It could be appropriate to say this of
  the circumstance concerning \persd; but the circumstance
  concerning \persc, \eg, can hardly be called a `cause'
  or `mechanism': it is only out of sheer ignorance that
  we assign, conditionally upon it, the distribution
  $(1/2, 1/2)$. Here and in the following, `circumstance'
  will generally mean simply what its name denotes: `a
  possibly unessential or secondary condition, detail,
  part, state of affair, factor, accompaniment, or
  attribute, in respect of time, place, manner, agent,
  \etc, that accompanies, surrounds, or possibly
  determines, modifies, or influences a fact or event'
  (\cf~\citep{randomhouse2003}).
\end{rem}

\section{Grouping the circumstances in a special way}
\label{sec:group_circ_spec_way}

The crucial step now is the following. Suppose that these
four circumstances interest us not for their intrinsic
details, but only in connexion with the plausibility
distributions they lead to for the coin toss
in the context $\zIc$.  In this regard, the circumstance
\prop{\persa\ tossed the coin}\footnote{Our knowledge
  about \persa\ must also be understood as implicit in
  this sentence; otherwise we should write \prop{\persa,
    who is a magician \etc, tossed the coin}. This also
  holds for the sentences that follow.}  and the
circumstance \prop{\persd\ tossed the coin} are for us
\emph{equivalent}, since both lead to the plausibility
distribution $(1, 0)$, as shown by \eqns~\eqref{eq:persa}
and~\eqref{eq:persd}. Similarly, \prop{\persb\ tossed the
  coin} and \prop{\persc\ tossed the coin} are also
equivalent, both leading to $(1/2, 1/2)$; \cf\
\eqns~\eqref{eq:persb} and~\eqref{eq:persc}. We should
like to have a set of circumstances such that different
circumstances led to different distributions.  The
first thing that comes to mind is to take the set
$\set{\zSpa \lor \zSpd, \zSpb \lor \zSpc}$ of the
disjunctions of equivalent circumstances, \ie\ $\zSpa
\lor \zSpd \equiv{}$\prop{\persa\ or \persd\ tossed the
  coin} and $\zSpb \lor \zSpc \equiv{}$\prop{\persb\ or
  \persc\ tossed the coin}. We must see, however, whether
this `coarse-grained' set really fulfils our wishes.

A simple theorem of plausibility theory comes to
help. It says that, in a given context, the
plausibility of a statement $A$ conditional on a
disjunction of mutually exclusive propositions $\set{B_j}$
is given by a convex sum of the plausibilities conditional
on the single propositions, as
follows~\citep[\chap~2]{jaynes1994_r2003}:
\begin{equation}
\label{eq:general_formula} \pr[A \cond
  ({\textstyle\Lor}_j B_j) \land \zI]= \sum_j \pr(A \cond
  B_j \land \zI) \frac{\pr(B_j \cond \zI) } {\sum_l
    \pr(B_l \cond \zI) }
\quad
  \text{($\set{B_j}$ mutually exclusive),}
\end{equation}
the weights being proportional to the plausibilities of
the $\set{B_j}$.  Note that the value of the plausibility
conditional on the disjunction, $\pr[A \cond
({\textstyle\Lor}_j B_j) \land \zI]$, generally depends on
the values of the plausibilities of the $\set{B_j}$,
$\set{\pr(B_j \cond \zI)}$. Thus, the latter plausibilities
must in general be specified if we want to find the first,
and that varies as these vary.  However, we see that this
dependence disappears when the plausibilities conditional
on each single $B_j$, $\set{\pr(A \cond B_j \land \zI)}$,
have all the same value (the right-hand side becomes a
convex sum of identical points). In this case also the
plausibility conditional on the disjunction, $\pr[A \cond
({\textstyle\Lor}_j B_j) \land \zI]$, will have that same
value, \emph{irrespective of the plausibilities of the
  $\set{B_j}$}:\footnote{Cases of vanishing plausibilities
  can be treated as appropriate limits. One can adopt the
  consistent convention that the product of an undefined
  plausibility (such as those with a contradictory
  context) times a defined and vanishing one also vanishes.}
\begin{multline} \label{eq:general_theorem}
\text{if }
\pr(A \cond B_j \land \zI)= \zq\text{ for all $j$},
\quad
\text{ then }
\pr[A \cond ({\textstyle\Lor}_j B_j) \land \zI]=\zq,
\\
\text{regardless of the values of the $\set{\pr(B_j \cond \zI)}$.}
\end{multline}

Clearly this is just the case when $A$ is either of our
outcomes $\set{\zR_i}$ and the $\set{B_j}$ are either pair
of equivalent circumstances. In fact, the protasis of
the last formula is just our previous definition of
\emph{equivalence} amongst circumstances.
Hence, the plausibility distribution for the results
conditional on the disjunction $\zSpa \lor \zSpd$ is the
same as those conditional on the two disjuncts separately,
\begin{equation}  \label{eq:plausfromdisjAD}
    \Bigl(\pr[\zR_i \cond (\zSpa \lor \zSpd)\land \zIc]
\Bigr) =
\Bigl(\pr(\zR_i \cond \zSpa \land \zIc) \Bigr) =
\Bigl(\pr(\zR_i \cond \zSpd \land \zIc) \Bigr) =
(1,0),
\end{equation}
and analogously for $\zSpb \lor \zSpc$:
\begin{equation} \label{eq:plausfromdisjBC}
  \Bigl(\pr[\zR_i \cond (\zSpb \lor \zSpc)\land \zIc]
  \Bigr) =
  \Bigl(\pr(\zR_i \cond \zSpb \land \zIc) \Bigr) =
  \Bigl(\pr(\zR_i \cond \zSpc \land \zIc) \Bigr) =
  \Bigl(\tfrac{1}{2},\tfrac{1}{2} \Bigr).
\end{equation}
This is true whatever the plausibilities of our four
initial circumstances might be (in fact, we have not yet
specified them!).

The coarse-grained set $\set{\zSpa \lor \zSpd, \zSpb \lor
  \zSpc}$ has thus, \emph{by construction}, the special
feature we looked for: different circumstances lead to
different plausibility distributions for the outcomes. The
circumstances can therefore be \emph{uniquely indexed} by
the respectively assigned plausibility distributions, and
we denote them accordingly:
\begin{align} 
  \zS_{(1, 0)} &\defd \zSpa \lor \zSpd, \label{eq:circA}
&
  \zS_{\bigl(\frac{1}{2}, \frac{1}{2}\bigr)} &\defd \zSpb
  \lor \zSpc,
\end{align}
and call them \emph{plausibility-indexed circumstances}.
With this indexing system, and denoting $\zqq \defd
(\zq_\zih, \zq_\zit)$, the conditional plausibilities of
the outcomes can be written
\begin{equation}
  \label{eq:compactpop}
  \pr(\zR_i \cond \zS_{\zqq} \land \zIc) = \zq_i.
\end{equation}

The last expression is in many ways similar to that
defining the generalised Bernoulli
model~\eqref{eq:bernoulli_general}. Indeed, one of our
main points is the following: \emph{plausibility-like
  parameters used as arguments of plausibilities can always
  be interpreted 
  to stand for appropriate plausibility-indexed
  circumstances}.

In view of \eqn~\eqref{eq:compactpop}, someone could
interpret the symbol `$\zSq$' as \prop{The plausibility
  distribution for the $\set{\zR_i}$ is $\zqq$} (similarly
to the symbol `$A_p$' introduced by
Jaynes~\citep[lect.~18]{jaynes1954}\citep[lect.~5]{jaynes1959}\citep[\chap~18]{jaynes1994_r2003}).
But such an interpretation is obviously wrong.  Let us make
this point clear. The symbol `$\zS_{(1, 0)}$', \eg, stands
for \prop{\persa\ or \persd\ tossed the coin}, as
\eqn~\eqref{eq:circA} shows; and this proposition does not
concern plausibilities at all.  It is true that this
proposition is the \emph{only} one leading us to assign
the distribution $(1, 0)$; but it is so just because of a
trick, \viz\ the fact that we have \emph{grouped} and
\emph{indexed} the initial circumstances in a particular
way.  Borrowing some terminology from
logic,
we can say that the correspondence between the proposition
\prop{\persa\ or \persd\ tossed the coin} and the
distribution $(1, 0)$ is \emph{only a trick within the
  metalanguage} of our theory~\citep{copi1954_r1979,church1956_r1970,fitzpatrick1966,ebbinghausetal1978_t1984,hofstadter1979_r1999}.

\begin{rem}\label{rem:meta_vs_in}
  The use of statements like \prop{The plausibility of $A$
    is $p$} or \prop{Data are drawn from a distribution
    $f$} is universal. Of course, they can be simply
  interpreted as `Look, the context and the circumstance
  are such that the plausibility of $A$ (the data) is $p$
  ($f$)', and this can be enough for our purposes: we may
  not need to know all the details of the context and the
  circumstance.  But note that those statements are more
  precisely \emph{meta}statements, statements \emph{about}
  plausibility assignments. As in logic, the use of such
  kind of statements \emph{as arguments of} plausibility
  formulae is preferably avoided. First, because such
  statements usually make poor contexts. Compare the
  statements \prop{Either \persc, who is a skilled coin
    tosser with a predilection for `head', or \persd, who
    has a two-headed coin, tossed the coin} with \prop{The
    plausibility distribution for `head' and `tail' is
    $(1, 0)$}: the former gives some clues as to \emph{the
    grounds on which} the distribution $(1, 0)$ is
  assigned, whereas the latter says only \emph{that} that
  distribution is assigned.\footnote{It reminds of
    Bachelierus' oft quoted answer: ``\langlatin{Mihi a
      docto Doctore/ Domandatur causam et rationem, quare/
      Opium facit dormire?/ \`A quoi respondeo,/ Quia est
      in eo/ Virtus dormitiva./ Cujus est natura/ Sensus
      assoupire}''~\citep[\french{troisi\`eme
      interm\`ede}]{moliere1673_r1682}.} Second, because
  such statements used inside plausibility formulae may
  give rise to self-references, circularity, and thus
  known paradoxes (`This proposition is false') and
  other
  inconsistencies~\citep{hofstadter1979_r1999,barwiseetal1987}.\footnote{\label{fn:expect}We
    find an example in an article by Friedman and
    Shimony~\citep{friedmanetal1971}. They introduce a
    proposition which says that the expectation of a
    certain quantity has a given value (their eq.~(4)).
    But such a proposition is a metastatement, because
    \emph{expectation} is defined in terms of plausibility
    assignments (in contrast to \emph{average}, which is
    defined in terms of measured
    frequencies~\citep{iso1993b}). The authors, however,
    do not notice this and proceed to use that proposition
    inside plausibilities, obtaining peculiar
    conclusions. Gage and Hestenes~\citep{gageetal1973}
    apparently show that these conclusions are not
    inconsistent, although they do not notice the mix-up
    of language and metalanguage either.
    Cyranski~\citep{cyranski1978} has a partially clearer
    view of the matter. \Cf\
    remark~\ref{rem:gouping_exp_value}. A metastatement
    inside a plausibility is used, although tentatively,
    also by Jaynes (his
    `$A_p$')~\citep[lect.~18]{jaynes1954}\citep[lect.~5]{jaynes1959}\citep[\chap~18]{jaynes1994_r2003};
    but our analysis shows that his ideas can be realised
    without this artifice.}
\end{rem}

\section{Analysis by marginalisation}
\label{sec:analys_marginal}

Let us now introduce the plausibilities of the original
circumstances in the context $\zIc$. For concreteness we
can assume them to be equally plausible:
\begin{equation}
  \pr(\zSpa \cond \zIc) 
= \pr(\zSpb  \cond \zIc) 
= \pr(\zSpc  \cond \zIc) 
= \pr(\zSpd \cond \zIc) = \frac{1}{4}.\label{eq:circumprior}
\end{equation}
From these values and the definitions~\eqref{eq:circA} we
have by the sum rule the plausibilities of the
plausibility-indexed circumstances:
\begin{align}
  \label{eq:priorcomp}
\pr[\zS_{(1,0)}\cond \zIc] &= \frac{1}{2},
&
\pr[\zS_{\bigl(\frac{1}{2}, \frac{1}{2}\bigr)}\cond \zIc] 
&= \frac{1}{2}.
\end{align}

These plausibilities can be used to write the distribution
for the outcomes on context $\zIc$ by marginalisation
over the circumstances. We can do this both with the
initial set $\set{\zSp_j}$ and with the set of
plausibility-indexed set $\set{\zSq}$. With the
first we obtain
\begin{equation}
  \label{eq:decompcoin_initial}
  \Bigl(\pr(\zR_i \cond \zIc)\Bigr) = 
\sum_j  
\Bigl(\pr(\zR_i \cond \zSp_j \land  \zIc)\Bigr) \,  
\pr(\zSp_j \cond  \zIc) 
= \Bigl(\tfrac{3}{4},\tfrac{1}{4} \Bigr).
\end{equation}
With the second set we must of course obtain,
consistently, the same result; but the decomposition has a
more suggestive (and possibly misleading!)\ form:
\begin{equation}
  \label{eq:decompcoin}
  \begin{aligned}
    \Bigl(\pr(\zR_i \cond \zIc)\Bigr) &=
    \sum_{\zqq} \Bigl(\pr(\zR_i \cond \zSq \land
    \zIc)\Bigr) \,
    \pr(\zSq \cond \zIc),
    \\
&=    \sum_{\zqq}
    \zqq \, \pr(\zS_{\zqq} \cond \zIc) =
    \Bigl(\tfrac{3}{4},\tfrac{1}{4} \Bigr) .
  \end{aligned}
\end{equation}
The index $\zqq$ assumes the two values $\set{(1, 0),\,
  (1/2, 1/2)}$, but we can let it range over the whole
simplex $\zrA$ defined in~\eqref{eq:simplex_def},
introducing a density function $\zqq \mapsto \zp(\zqq
\cond \zIc)$ in the usual way (explained later in
\sect~\ref{sec:gener_rem}). In this case it is given
by
\begin{equation}\label{eq:deltabernoulli}
\zp(\zqq \cond \zI)\defd
  \frac{1}{2}\,\Bigl[\delt(\zq_\text{h} -1)
+
\delt\Bigl(\zq_\text{h} - \tfrac{1}{2}
\Bigr)\Bigr]\, \delt(\zq_\text{h} +\zq_\text{t} -1),
\end{equation}
a weighted sum of Dirac deltas\footnote{In Egorov's
  sense~\citep{egorov1990,egorov1990b,demidov2001}; see
  also~\citep{lighthill1958_r1964,delcroixetal2002,delcroixetal2004,oberguggenberger2001}
  and \cf~\citep{swartz2001,bartle2001,pfeffer1993}.} with support on
$\zqq=(1,0)$ and $(1/2, 1/2)$. The marginalisation~\eqref{eq:decompcoin}
thus takes the form
\begin{equation}\label{eq:margincoin}
  \Bigl(\pr(\zR_i \cond \zIc)\Bigr)= 
  \int_\zrA \zqq\, \zp(\zqq\cond \zIc)\, \di\zqq,
\end{equation}
which is similar to the formula~\eqref{eq:marginbernoulli}
for the generalised Bernoulli model.

\section{Updating the plausibility of the circumstances}
\label{sec:updating_ex1}

If the outcome of the toss is, say, `head', what do the
plausibilities of the circumstances become? In other
words, what are the circumstances' plausibilities in the
context $\zRH \land \zIc$? The answer is obviously given
by Bayes' theorem:
\begin{equation}
  \label{eq:update_head_ex1}
  \pr(\zSq \cond \zRH \land \zIc)
=
\frac{
\pr(\zRH \cond \zSq \land \zIc)\,
\pr(\zSq \cond \zIc)
}{
\sum_{\zqq'}
\pr(\zRH \cond \zSqp \land \zIc)\,
\pr(\zSqp \cond \zIc)
},
\end{equation}
or, in terms of the density $\zp$,
\begin{multline}
  \label{eq:update_head_ex1_pl_dens}
  \zp(\zqq \cond \zRH \land \zIc)
=
  \frac{ \zqh\, \zp(\zqq \cond  \land\zIc) }
{ \int_\zrA \zqh'\, \zp(\zqq' \cond \land\zIc)
    \,\di\zqq' }
={}
\\
  \biggl[\frac{2}{3}\,\delt(\zq_\text{h} -1)
+
\frac{1}{3}\,\delt\Bigl(\zq_\text{h} - \tfrac{1}{2}
\Bigr)\biggr]\, \delt(\zq_\text{h} +\zq_\text{t} -1).
\end{multline}
The plausibility of $\zS_{(1,0)}$, \ie, that \persa\ or
\persd\ tossed the coin, has thus increased a little.

\begin{rem}\label{rem:no_dist_equiv_updating}
  Note that knowledge of the outcome can help to increase
  the plausibility of one of the plausibility-indexed
  circumstances $\set{\zSq}$ at the expense of the
  others', but can never do so within a set of
  \emph{equivalent} circumstances like $\set{\zSpa,
    \zSpd}$ or $\set{\zSpb, \zSpc}$.
\end{rem}

\medskip

The last formula is a very simple instance of the answer to
an inverse problem.  Our point is, again, that the
marginalisation over a plausibility-like parameter and the
updating of its distribution can be interpreted as the
same operations for a set of plausibility-indexed
circumstances.  From this standpoint, and as should be
clear from a previous discussion and remarks, the
plausibility $\pr(\zSq \cond \zIc)$ (and its density
$\zp(\zqq \cond \zIc)$) is not the plausibility of a
plausibility, but simply the plausibility of a
circumstance, the latter being indexed in a particular
way.

\chapter{Second Example: multiple measurements, particular
  convex structures of circumstances, updating}
\label{sec:second_example}

\section{Context}
\label{sec:ex2_context}

The following example differs from the first in the number
of measurements and circumstances considered. Consequences:
the space of parameters has particular convex structures,
and the plausibilities of the outcomes of one measurement
can be `updated' upon knowledge of the outcome of the
other.

We have a box with two buttons, marked `Letter' and
`Number', and a display. Push the `Letter' button, and
either `a' or `b' appears on the display; push `Number',
and `1' or `2' appears. We can push each button only once,
and only one at a time. Call, improperly,
\emph{measurement} the act of pushing a button and reading
the display; call `outcome' what is then read on the
display. Denote the `Letter' measurement by $\zMl$ and its
outcomes by $\set{\zRa, \zRb}$; the `Number' measurement by
$\zMn$ and its outcomes by $\set{\zRo, \zRt}$.

Given only the above knowledge, we should assign a
plausibility distribution $(1/2, 1/2)$ to the outcomes of
each measurement. But we know in fact something more about
the construction of the box: inside, besides some sort of
machinery, there is a chest containing an even number, $2
N$, of balls. Each ball is marked either `a1',
`a2', `b1', or `b2'. When a button is pushed, the
machinery draws one of the balls from the chest and sends,
depending on the button, either the letter \emph{or} the
number printed on the drawn ball to the display; and then
puts the ball back into the chest.\footnote{This renders
  the temporal order of the measurements (if both are
  performed) irrelevant. That is why we are not making
  temporal considerations. (Note also that we do not need
  to suppose that the urn is shaken after the replacement
  of the ball: this would add nothing to our state of
  knowledge, since we do not know how the machine makes
  the replacement anyhow.)} We have also a very important
piece of knowledge as to how the $2 N$ balls were
originally chosen and put into the chest: this initially
contained $4 N$ balls, marked `a1', `a2', `b1', and `b2'
in equal proportions (\ie, $N$ balls marked `a1', $N$
`a2', \etc). From these, $2 N$ balls where taken away, so
only $2 N$ remained in the chest.  This is all we know;
denote it (together with everyday knowledge concerning
balls, buttons, boxes, \etc)\ by $\zIn$.

From $\zIn$, some points are immediately clear. First, not
all the $2 N$ balls in the chest can be marked `a1', nor
all `a2', \etc, since the chest initially contained only
$N$ of each type. Second, if all the $2 N$ balls have the
`a' mark, then $N$ of them must necessarily be of the `a1'
kind and the other $N$ must be of the `a2' kind. Similarly
for the marks `b' and, exchanging the r\^ole of letters and
numbers, `1' and `2'.

\section{Introducing a set of circumstances}
\label{sec:intro_circ_ex2}

Let us analyse the context $\zIn$ into a set of mutually
exclusive and exhaustive possible circumstances. Different
choices are possible. One is to consider the possible sets
of balls left in (or equivalently, taken away from) the
chest.  The number of circumstances thus defined is given
by the number of ways of choosing $2 N$ objects from a
collection of $4 N$ distinct ones without regard to order
--- the binomial coefficient $\tbinom{4 N}{2 N}$. Note
that it matters \emph{which} of the `a1'-marked balls are
chosen, and likewise for the others.  Our knowledge is
symmetric in respect of these circumstances, hence they
are assigned equal plausibilities.\footnote{That is, they
  are assigned equal plausibilities not because `the
  balls are initially chosen at random' or something of
  the kind, but because we just do not know how they have
  been chosen. In fact, they can have been chosen
  according to a particular scheme; the point is that we
  do not know such scheme.}

Another choice is to consider as a circumstance the
numbers of balls marked `a1', `a2', \etc\ left in the
chest instead. Note the difference with the previous
choice: this is a sort of `coarse graining' thereof. For
this reason the newly defined circumstances are not
equally plausible. We settle for this second choice and
denote a generic circumstance by
$\zSp_{\alpha\text{a1},\beta\text{a2},\gamma\text{b1},\delta\text{b2}}$,
meaning \prop{$\alpha$ `a1'-marked balls, \ldots, and
  $\delta$ `b2'-marked balls are left in the chest}. The
coefficients $\alpha$, $\beta$, \etc\ must obviously sum
up to $2N$ and each can range from $0$ to $N$.

\section{Plausibility-indexing the circumstances; their
particular set}
\label{sec:plaus_ind_circ_ex2}

As in the previous example, suppose that we are not
interested in the details of the circumstances above, but
only in the plausibilities they lead us to assign to the
outcomes of the two measurements $\zMl$ and $\zMn$. We can
group the circumstances into plausibility-indexed
equivalence classes, as before. In the present case the
equivalence must take into account \emph{two} plausibility
distributions, one for each measurement.

Here is an example for $N=2$. The two different
circumstances $\zSpexa$ and $\zSpexb$ lead both to the
same plausibility distribution $(1/2, 1/2)$ for the
`Letter' measurement, and to the same distribution $(3/4,
1/4)$ for the `Number' measurement (as is clear by simply
counting their `a's, `b's, `1's, and `2's).  Moreover,
\emph{only} these two circumstances lead to the
plausibility distributions above, as the reader can prove. By
theorem~\eqref{eq:general_theorem}, also their disjunction
$\zSpexa \lor \zSpexb$ leads to the same distributions and
can thus be denoted by
\begin{equation}
  \label{eq:composite_spex_ex}
  \zSexab \defd \zSpexa \lor \zSpexb.
\end{equation}
This is one of the plausibility-indexed circumstances. Its
plausibility is the sum of its disjuncts' plausibilities,
$\pr[\zSexab \cond \zIn] = \pr(\zSpexa \cond \zIn) +
\pr(\zSpexb \cond \zIn)$.

\begin{figure}[pbtr]
  \begin{center}
    \includegraphics[width=0.49\textwidth]{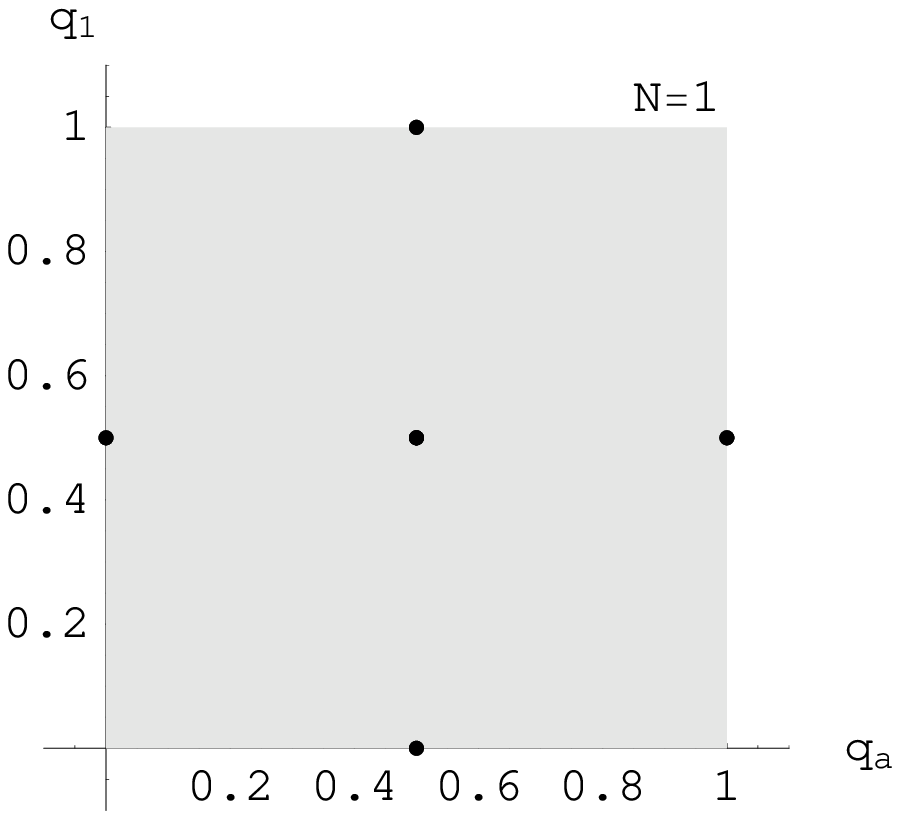}
    \includegraphics[width=0.49\textwidth]{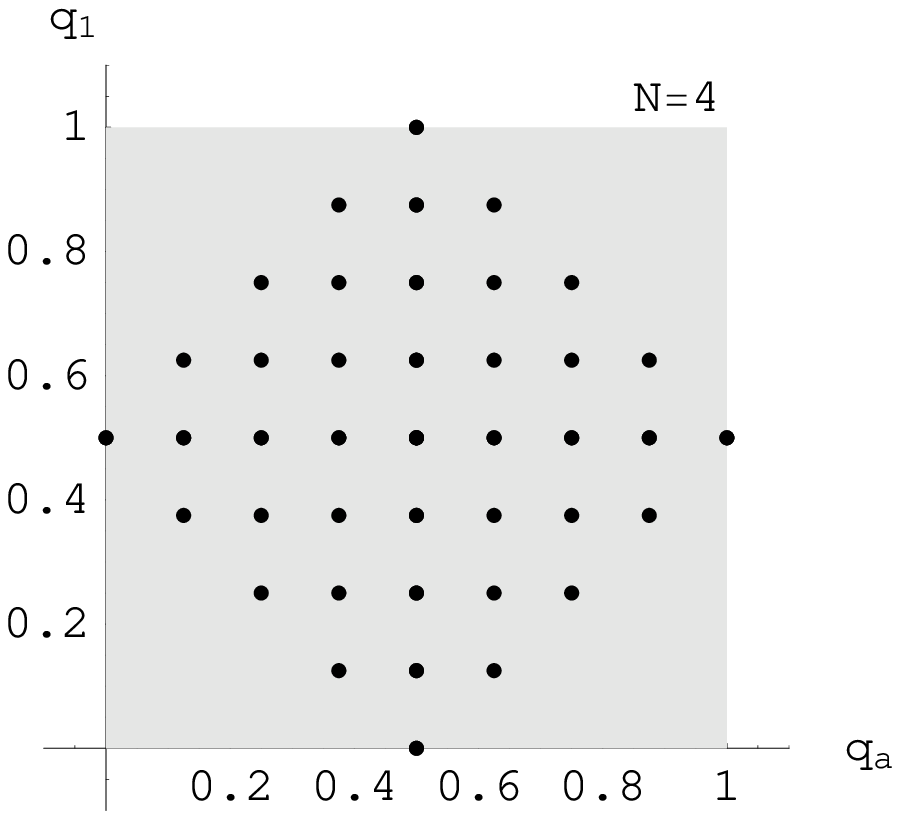}
    \\
    \includegraphics[width=0.49\textwidth]{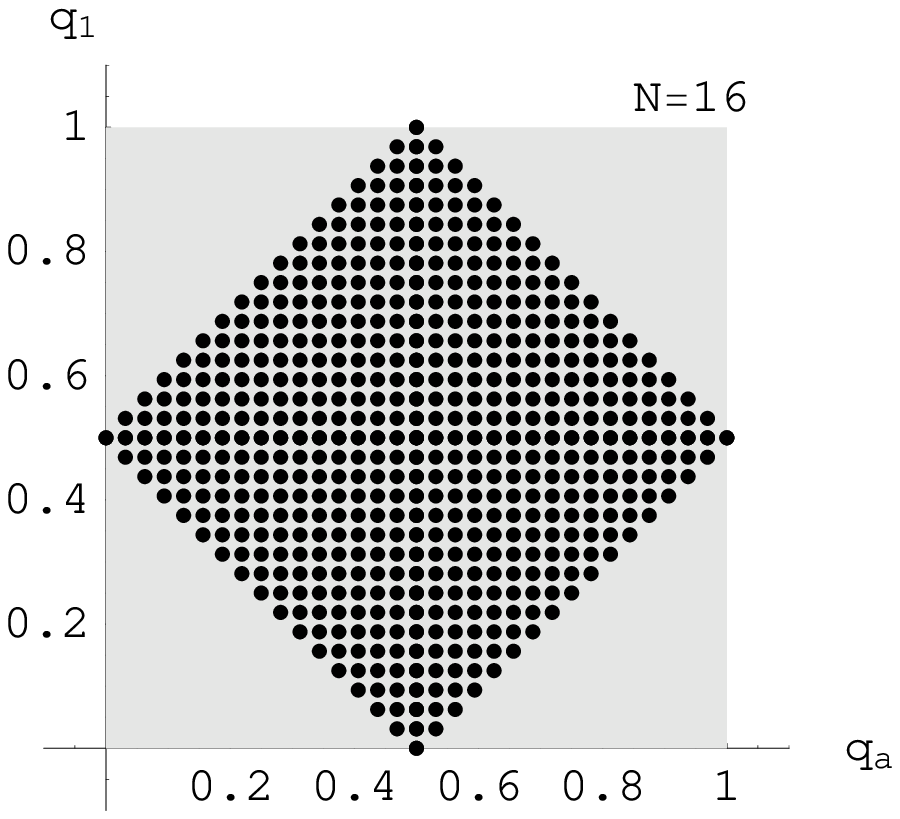}
  \end{center}
\caption{
Possible values of the plausibilities
  $\zq^\text{L}_\text{a}$ and $\zq^\text{N}_1$ for $N=1$,
  $4$, and $16$. They are in bijective correspondence with
  the plausibility-indexed circumstances. The grey region
  is the set of all possible pairs of plausibility
  distributions for two generic
  measurements.\label{fig:n1}}
\end{figure}

In general, for any $N$, we have plausibility-indexed
circumstances denoted by $\zSql$, the parameters $\zqql$
and $\zqqn$ corresponding to the plausibility
distributions for the `Letter' and the `Number'
measurements. The indexing is such that
\begin{equation}
  \label{eq:ball_prop_Sq}
  \pr(\zR^k_i \cond \zM^k \land \zSql \land \zIn) =
\zq^k_i
\qquad
\text{for $k=\text{L}, \text{N}$ and all appropriate $i$.}
\end{equation}
We leave to the reader the pleasure of proving that there
is a total of $N^2 + (N+1)^2$ plausibility-indexed
circumstances, \ie, of distinct values for the parameters
$(\zqql,\zqqn)$. They can be represented by points on the
plane $\zq^\text{L}_\text{a} \zq^\text{N}_1$ as
illustrated in fig.~\ref{fig:n1} for the cases $N=1$,
$N=4$, and $N=16$ respectively. It is not difficult to see
(especially looking at the figure for $N = 16$) that as
$N\to \infty$ their set $\zQE$ becomes dense in the convex
set $\zQQE$ defined by
\begin{equation}
  \label{eq:rhombus}
  \zQQE \defd 
\set{(\zqql,\zqqn) \st  \znorm{\zqql} +   \znorm{\zqqn} \le 1},
\end{equation}
where $\znorm{\zqq}$ is the supremum norm $\znorm{\zqq}
\defd \max_i\set{\zq_i}$. Thus in the limit $N\to \infty$ we
may effectively work with a continuum of
plausibility-indexed circumstances in bijection with the
points of this set. Denote this `limit context' by
$\zIi$.

We observe two interesting facts. The first is that $\zQQE$
is a \emph{proper} subset of the set of all possible pairs
of plausibility distributions for two generic measurements
(the grey square region in the figure); the latter is the
Cartesian product of two one-dimensional simplices,
${\zrA_1} \times {\zrA_1}$. We could have let the
parameters $(\zqql,\zqqn)$ range over the latter set; in
this case, however, the contexts $\zIn$ and $\zIi$ would
have led us to a vanishing plausibility density for those
parameter values not belonging to $\zQQE$.
The second interesting fact is that neither $\zQQE$ nor the
larger set ${\zrA_1} \times {\zrA_1}$ are simplices. It is
so because we are considering \emph{two} measurements (had
we considered a single measurement with four outcomes, we
would have dealt with a three-dimensional simplex
instead).%
\footnote{You might ask: ``Couldn't we consider a single
  measurement with the four outcomes `a1', `a2', `b1', `b2'
  instead?''. The answer is: yes, we could have introduced
  a single \emph{fictive} measurement with $\zMl$ and
  $\zMn$ arising as marginals. But what for?  After all,
  the rules of this game do not make allowance for such a
   measurement.} See also remark~\ref{rem:phys_th_sets}.

\section{Analysis by marginalisation}
\label{sec:ex2_analys_margin}

We can write the plausibility distributions for the
measurements as marginalisations over the
plausibility-indexed circumstances $\set{\zSql}$, using
the latter's plausibilities $\set{\pr[\zSql \cond \zIn]}$.
Denote for brevity $(\zqql,\zqqn)\defs \zqqq$, hence
$\set{\zSql}\equiv\set{\zSqq}$. Then
\begin{multline}
  \label{eq:marginal_balls}
  \begin{aligned}
    \Bigl( \pr(\zR^k_i \cond \zM^k \land \zIn) \Bigr) &=
    \sum_{\zqqq \in \zQE} \Bigl( \pr(\zR^k_i \cond \zM^k
    \land \zSqq \land \zIn) \Bigr)\, \pr(\zSqq \cond \zIn),
    \\
&=    \sum_{\zqqq \in \zQE} \zqq^k\, \pr(\zSqq \cond \zIn),
  \end{aligned}
\\
k=\text{L},\text{N}
\end{multline}
Also this formula, like~\eqref{eq:marginbernoulli} and
\eqref{eq:decompcoin}, looks like a weighted sum of
plausibilities, and $\pr(\zSqq \cond \zIn)$ looks like
the plausibility of two plausibility distributions. But
this is not the case, just as it was not in the example of
the coin: the propositions $\set{\zSqq}$ speak not about
plausibilities but about possible preparations of the box
and its contents; yet they are suitably indexed according
to the plausibilities they lead us to assign to the
measurements' outcomes.

The sums above can also be replaced by integrals over the set
$\zQQE$,
\begin{gather}
  \label{eq:marginal_balls_infty}
  \Bigl(
\pr(\zR^k_i \cond \zM^k \land \zIn)
\Bigr)
=
\int_\zQQE 
\zqq^k\,
\zp(\zqqq\cond \zIn)\,\di\zqqq,
\end{gather}
where the density $\zqqq \mapsto \zp(\zqqq\cond \zIi)$ is
introduced just like in the example of the coin.

\section{`Updating' the plausibilities of the
  circumstances and the outcomes}
\label{sec:bayes_circ}

Suppose that the `Letter' button has been pushed and the
outcome `a' has appeared on the display. This knowledge
places us in a new context, expressed by the proposition
$\zMl \land \zRa \land \zI$. We ask: (1) What
plausibilities
\begin{equation}
  \label{eq:plau_Sq_new}
  \pr(\zSqq \cond \zRa \land \zMl \land \zIn)
\end{equation}
do we assign to the plausibility-indexed circumstances
$\set{\zSql}$ in the new context? Furthermore, we still
have the possibility of pushing the `Number' button
once. So we also ask: (2) What plausibilities
\begin{equation}
  \label{eq:plaus_Ri_from_Ra}
  \pr(\zRnu_i \cond \zMn \land
  \zRa \land \zMl \land \zIn)
\end{equation}
do we now assign to the outcomes $\set{\zRo, \zRt}$ in case we
push the `Number' button?

Let us answer the first question.  We use the assumption
(valid in the context $\zIn$) that
knowledge of the performance of any of the two measurements
(but not of their outcomes!)\ is irrelevant for assigning
plausibilities to the circumstances:
\begin{equation}
  \label{eq:ass_1_Ri}
  \pr(\zSqq \cond \zM^k \land \zIn)
  =
  \pr(\zSqq \cond \zMl \land
  \zMn \land \zIn)
   =
  \pr(\zSqq \cond \zIn)
\qquad
\text{for all $\zqqq$, $k$}.
\end{equation}
With this assumption and \eqn~\eqref{eq:ball_prop_Sq},
Bayes' theorem yields a simplified form for the sought
plausibilities~\eqref{eq:plau_Sq_new}:
\begin{equation}
  \label{eq:plau_Sq_from_Ra}
  \pr(\zSqq \cond \zRa \land \zMl \land \zIn) 
=
  \frac{\zq^\text{L}_\text{a}\, \pr(\zSqq \cond \zIn)}
  {\sumqql {{\zq'}^\text{L}_\text{a}}\, \pr(\zSqql \cond
    \zIn)}.
\end{equation}
This is also the answer to an inverse problem. Note, again,
that it expresses the updated plausibility distribution,
not `of the parameter $\zqqq$', but of propositions like
\prop{There are $\alpha$ `a1'-marked balls, \ldots, and
  $\delta$ `b2'-marked balls left in the chest; or \ldots;
  or $\alpha'$ `a1'-marked balls, \ldots, and $\delta$
  `b2'-marked balls left in the chest}.

To answer the second question we use, beside
assumption~\eqref{eq:ass_1_Ri}, the following fact, which
holds in our context $\zIn$: If we want to determine the
plausibility distribution for one of the measurements, and
we know which particular circumstance holds, then for us
it is irrelevant to know whether the other measurement has
been performed, or which outcome it has yielded. For
example, if we are interested in the plausibilities of the
outcomes of the `Number' measurement, and we know that a
particular circumstance $\zSqq$ holds (\eg, that in the
chest there are two `a1'-marked balls, one `a2'-marked
ball, \etc; or one `a1'-marked ball, one `a2'-marked ball,
\etc), then knowledge of the outcome of the mere
performance of `Letter' measurement is irrelevant.  In
formulae,
\begin{multline}
  \label{eq:ass_2_Ri}
  \pr[\zR^\ka_i \cond (\zR^\kb_j \land\zM^\kb) \land
  \zM^\ka \land \zSqq \land \zIn]
=
  \pr(\zR^\ka_i \cond \zM^\kb \land
  \zM^\ka \land \zSqq \land \zIn)
  ={}
\\
  \pr(\zR^\ka_i \cond \zM^\ka \land \zSqq \land \zIn)
\quad \text{for all $\ka$, $\kb\neq\ka$, and appropriate $i$, $j$}.
\end{multline}
Analysing \eqn~\eqref{eq:plaus_Ri_from_Ra} in terms of
circumstances and using \eqns~\eqref{eq:plau_Sq_from_Ra}
and~\eqref{eq:ass_2_Ri} we find
\begin{equation}
  \label{eq:plaus_Ri_from_Ra_analysed}
  \pr(\zRnu_i \cond \zMn \land
  \zRa \land \zMl \land \zIn)
=
\frac{\sumqq \zq^\text{N}_i\,
\zq^\text{L}_\text{a}
\,
 \pr(\zSqq \cond \zIn)}
{\sumqql
{\zq'}^\text{L}_\text{a}
\,
 \pr(\zSqql \cond \zIn)}.
\end{equation}

In regard to the assumptions summarised in
\eqns~\eqref{eq:ass_1_Ri} and~\eqref{eq:ass_2_Ri}, \cf\
remark~\ref{rem:assumptions_notalways_hold}.

\chapter{Generalisation and summary of principal formulae}
\label{sec:gener_rem}

The two examples should suffice to give an idea of the
interpretation of $\zqqq$-like parameters and of their
plausibilities, and of the principal consequences of this
interpretation. The reader could try to make similar
analyses for the toy models by
Kirkpatrick~\citep{kirkpatrick2002b,kirkpatrick2001,kirkpatrick2002},
Spekkens~\citep{spekkens2004}, or
us~\citep{mana2004,mana2004b}.  We shall now present the
idea in general and abstract terms. Some additional
remarks will also be given.

\section{Experiments, outcomes, circumstances}
\label{sec:exp_out_circ}

In the general case we have a context $\zI$ and a set of
$\zm$ \emph{measurements}, represented by propositions
$\zM^k$, $k=1,\dotsc,\zm$. Each measurement $\zM^k$ has
mutually exclusive and exhaustive \emph{outcomes}
represented by a set of propositions $\set{\zR^k_i}$. The
number of outcomes can vary from measurement to measurement,
so that $i$ ranges over appropriate sets for different $k$.
The index $k$ is omitted when no confusion arises.

\begin{rem}\label{rem:termin_exp_out}
  The use of the terms `measurement' and `outcome' is only
  dictated by concreteness. The formalism and the
  discussion presented apply in fact
  to more general concepts. What we call `measurement'
  could be only a ca\emph{su}al observation, or simply a
  `state of affairs' which can present itself in mutually
  exclusive and exhaustive `forms' (the `outcomes'). The
  term `measurement' shall hence be divested here of those
  connotations implying active planning and control, which
  are not relevant to our study.
Moreover, a `measurement'
  needs not be associated with a point or short interval
  in time or space. It can \eg\ be a collection of
  observations; in this case its `outcomes' are all
  possible combinations of results from these
  observations.
  Finally, note that the $\zm$ measurements are generally
  different, \ie, they are not necessarily `repetitions' of
  the `same' measurement --- a case that will be
  discussed in the second paper instead.
\end{rem}

A set of \emph{circumstances} $\set{\zSp_j}$ is
introduced; these represent a sort of more detailed,
possible descriptions of the context $\zI$, and are mutually
exclusive and exhaustive, \ie\ we know that one and only
one of them holds:
\begin{gather}
  \label{eq:S_mut_excl_exh}
  \pr(\zSp_{j'} \land \zSp_{j''} \cond \zI) = 0
\quad\text{for all $j'$ and $j''\neq j'$},\\
\pr({\textstyle\Lor}_j \zSp_j \cond \zI) = 1.
\end{gather}

The plausibilities of the measurements' outcomes
conditional on the circumstances,
\begin{equation}
  \label{eq:all_plaus}
  \pr(\zR_i \cond \zM^k \land \zSp_j \land \zI)\quad
  \text{for
    all $j$, $k$, and appropriate $i$},
\end{equation}
are assumed to be given.

\begin{rem}\label{rem:circumst_rem_2}
  The notion of `circumstance', represented by
  propositions $\zSp_j$ and later also $\zS_{\zqq}$, has
  been further explained in
  remark~\ref{rem:circum_no_causes}.  An example of
  circumstance from \sect~\ref{sec:introexample} is
  \prop{\persb\ tossed the coin}; other examples are
  \prop{The temperature during the experiment was
    $25\,\uni{\celsius}$} and the more elaborated \prop{We
    studied the density of monodisperse spherical
    particles in a tall cylindrical tube as a series of
    external excitations, consisting of discrete, vertical
    shakes or `taps,' were applied to the
    container}~\citep{nowaketal1998}.
  As in the case of `measurement', a circumstance needs not
  be related to a single point or short interval in space
  or time. For example, in assigning the plausibility that
  it will rain or has rained in a given place at a given time, a
  circumstance might consists in a specific history of
  worldwide meteorological conditions under the preceding
  two years.
  For reasons discussed in
  remark~\ref{rem:meta_vs_in}, we require that a
  circumstance be described or specified in concrete
  terms, and metastatements like \prop{The samples are
    drawn from a distribution $f$} or \prop{The
    plausibility of head is $1/3$} are excluded.
%
  Finally, the choice of an appropriate set of
  circumstances, \ie, of the appropriate way and depth to
  analyse a particular problem (the context), can only be
  decided on an individual basis, of course.
\end{rem}

\section{Plausibility-indexing the circumstances}
\label{sec:plaus_index_gen}

The circumstances are then grouped into equivalence
classes. Two circumstances are equivalent if they lead to
the same plausibility distributions for each measurement
$\zM^k$:
\begin{equation}
  \label{eq:equiv_gener}
  \zSp_{j'} \equi \zSp_{j''} \iff{}
\left\{
  \begin{aligned}
    &\pr(\zR_i \cond \zM^k \land \zSp_{j'} \land
    \zI)  = \pr(\zR_i \cond \zM^k \land
    \zSp_{j''}
    \land \zI) \\
    &\text{for all $k$ and appropriate $i$}.
  \end{aligned}
\right.
\end{equation}

By construction the equivalence classes are in injective
correspondence with the possible numerical values of the
plausibility distributions for the measurements, $(\zqq^1,
\zqq^2, \dotsc)$. Denote a generic such value by $\zqqq \defd
(\zqq^k)$,
its equivalence class by $\zeqcl$, and membership by
$\zSp_j\in \zeqcl$ or simply $j \in \zeqcl$.
We take all disjunctions of equivalent
circumstances
\begin{equation}
  \label{eq:compos_def_gen}
  \zS_{\zqqq} \defd
\Lor_{j\in \zeqcl} \zSp_j,
\end{equation}
and call these (in lack of a better name)
\emph{plausibility-indexed circumstances}, shortened to
`circumstances' whenever no confusion is possible.
Conditional on such a circumstance $\zS_{\zqqq}$, the
plausibilities of the outcomes have numerical values
identical to its indices:
\begin{equation}
  \label{eq:com_circ_lead_q}
\pr(\zR^k_i \cond \zM^k \land
\zS_{\zqqq} \land \zI) 
= \zq^k_i,
\end{equation}
a formula that reminds of a generalised Bernoulli model
(\cf\ \eqn~\eqref{eq:bernoulli_general}).

Our main belief, already stated in the coin example, is
that \emph{plausibility-like parameters used as arguments
  of plausibilities can always be interpreted 
  to stand for some appropriate plausibility-indexed
  circumstances}.\footnote{\label{fn:diff_circ}What
  constitutes a circumstance is largely a matter of
  situation, purpose, and personal good taste as well. The
  formalism presented cannot think up the circumstances
  for us. In \sect~\ref{sec:introexample} we spoke \eg\
  about different persons' skills in coin-tossing; but
  other people could speak about different values of the
  coin's `propensity' to come up heads. Perhaps the
  reason why `de~Finettians' have always felt uneasy
  about plausibility-like parameters and their priors was
  that these mathematical objects leave room to ideas and
  concepts that are unnecessary or not in good taste (\cf\
  Jaynes~\citep{jaynes1994_r2003}, \chap~3, `Logic versus
  propensity'). To keep off these ideas they partially
  denied priors their meaning as plausibilities (this has
  led, fortunately, to some very beautiful ideas and
  theorems~\citep{definetti1937}). We hope to have shown
  here and in the next paper that there is no need to
  adopt such extreme measures.}

The passage to plausibility-grouped circumstances can have two main
motivations. (1) We can be interested in the plausibilities the
circumstances lead to, rather than in the latter's intrinsic details. (2)
We may want a set of circumstances with the property that knowledge of
outcomes can increase the plausibility of only one circumstance. This is
true for the set $\set{\zSqq}$, but not for the set $\set{\zSp_j}$ in
general. In fact, \emph{knowledge of new outcomes can never lead to a
  alteration of the ratios of the plausibilities of two or more equivalent
  circumstances.} \Cf\ remark~\ref{rem:no_dist_equiv_updating} and
see~\citep{portamana2007b,portamana2007}
and~\citep[\sect~5.3]{portamanaetal2007}.

\begin{rem}\label{rem:gouping_exp_value}
  Suppose that to each outcome $\zR_i$ of some measurement
  $\zM_k$ is 
  associated a value $\zr_i$ of some physical quantity, so
  that it makes sense to speak of the \emph{expected
    value}\footnote{Which should not be confused with the
    \emph{average}~\citep{iso1993b}, defined in terms of
    observed frequencies. \Cf\ footnote~\ref{fn:expect}.}
  of this quantity in a generic context $\zM_k \land \zJ$:
  \begin{subequations}
    \begin{equation}
      \label{eq:expected_def}
      \expe{\zr\cond \zM_k \land\zJ} \defd
      \sum_i \zr_i \, \pr(\zR_i \cond \zM_k \land \zJ).
    \end{equation}
    In our case, the formation of equivalence classes of circumstances can
    then be made with respect to expected values instead of plausibilities,
    \ie,
    \begin{equation}
      \zSp_{j'} \equi
      \zSp_{j''} \iff 
      \expe{\zr^k \cond \zM^k \land \zSp_{j'}
        \land \zI} = \expe{\zr^k \cond \zM^k \land \zSp_{j''}
        \land \zI} \text{ for all $k$.}
      \label{eq:expected_cons}
    \end{equation}
  \end{subequations}
In this way we obtain a set of
\emph{expectation-indexed} circumstances
$\set{\zS_{\overline{\expe{\zr}}}}$. Note that two different
circumstances in such a set (leading hence to different
expectations) may lead to the \emph{same}
probability distributions for the outcomes; therefore this set
is not to be confused with, and has not the same
applications of, our $\set{\zSqq}$.
\end{rem}

Particularly interesting is the space $\zQQ$ of the
parameters $\zqqq$. Since these correspond to numerical
values of plausibility distributions for the measurements,
$\zQQ$ is in general a (possibly proper) subset of a
Cartesian product of simplices $\varprod_k\zrA^{(k)}$, the
simplex $\zrA^{(k)}$ corresponding to the plausibility
distribution for the $k$th measurement.

\begin{rem}\label{rem:phys_th_sets}
  The features of the subset $\zQQ$ will depend on the
  nature of the circumstances $\set{\zSp_j}$ (and thus of
  the $\set{\zSqq}$). In some cases it is simply
  postulated that some kinds of circumstances do not
  present themselves, and this will delimit the subset
  accordingly. We saw an instance of this in the box
  example of \sect~\ref{sec:second_example}, in which the
  set $\zQQ$ was, for each $N$, a special proper subset
  $\zQE$ of the Cartesian product of two two-dimensional
  simplices (the grey square region in the figures).
  There are examples of physical theories where we
  postulate (by induction from numerous observations) that
  the set of `circumstances in which a system is
  prepared' --- often called \emph{states} --- is somehow
  restricted. This also restricts the space of the
  mathematical objects representing these states to
  particular, non-simplicial (convex) sets. The most
  notable example is quantum theory, in which the set of
  statistical operators --- the mathematical objects
  representing the states --- has very strange
  shapes~\citep{jakobczyketal2001,kimura2003,kimuraetal2004,portamana2006b}.\footnote{That is, if we represent this set so as
    to preserve its convex properties, which are the
    relevant ones (see the third note of this
    series).}
  The set of Gibbs distributions in classical statistical
  mechanics provides another example.
\end{rem}

\begin{rem}
  The plausibility-indexed circumstances need not be
  parametrised by the values of the plausibility
  distributions $(\zqq^k) \equiv \zqqq$. Other
  parametrisations can be used as long as they are in
  bijective correspondence with the $\zqqq$ one, and some
  may be more useful (\cf~\citep{mccullagh1992}). Usually,
  what is relevant is the convex structure of the set of
  parameters $\zQQ$, a point on which we shall return in
  the third paper.
\end{rem}

\section{Priors and analysis by marginalisation}
\label{sec:analys_marginal_gen}

If the initial circumstances $\set{\zSp_j}$ have the
plausibility distribution $\Bigl(\pr(\zSp_j \cond
\zI)\Bigr)$, by the sum rule the plausibility-indexed
circumstances have distributions
\begin{equation}
  \label{eq:distr_plau_ind}
  \pr(\zSqq \cond \zI) 
  = \sum_{j \in  \zeqcl} \pr(\zSp_j \cond \zI)
\end{equation}
(see also remark~\ref{rem:diff_priors_diff_circ}).

In terms of the plausibility-indexed circumstances, the
plausibility distribution for each measurement outcome
$\zR^k_i$ can be expressed in marginal form as
\begin{equation}
  \label{eq:decomp_general}
  \begin{aligned}
    \pr(\zR^k_i \cond \zM^k \land \zI) &= 
    \sum_{\zqqq\in\zQQ} \pr(\zR^k_i \cond \zM^k \land
    \zS_{\zqqq} \land \zI) \, \pr(\zS_{\zqqq} \cond \zI),
\\
&    \equiv
    \sum_{\zqqq \in \zQQ} \zq^k_i \, \pr(\zS_{\zqqq} \cond
    \zI), 
\\
&\equiv \int_{\zQQQ} \zq^k_i \, \zp(\zqqq \cond \zI)\, \di\zqqq,
  \end{aligned}
\end{equation}
(\cf\ \eqn~\eqref{eq:marginbernoulli}) where $\zqqq
\mapsto\zp(\zqqq \cond \zI)$ is an appropriate generalised
function~\citep{lighthill1958_r1964,colombeau1984,colombeau1985,colombeau1992}
(see also~\citep{bartle1996,gordon1998}).
The sudden appearance of an integral can be justified (as
customary) as follows: $\zqqq$ becomes a continuous
parameter whose range is some set $\zQQQ$ such that
$\convh\zQQ \subseteq \zQQQ \subseteq
\varprod_k\zrA^{(k)}$ (where $\convh\zQQ$ is the convex
hull of $\zQQ$), and we introduce a density function
$\zqqq \mapsto \zp(\zqqq\cond \zI)$ such that, for each
$\zrB\subseteq \zQQQ$ (from a suitable $\sigma$-field
of subsets), $\smallint_\zrB\,\zp(\zqqq\cond \zI)\,\di
\zqqq =\sum_{\zqqq' \in \zrB\cap\zQQ} \pr(\zSqql\cond
\zI)$.%
\footnote{No one forbids us to introduce additional
  \emph{impossible} fictitious circumstances
  $\set{\zSp_{j'}}$ (which may involve, \eg,
  `\emph{centaurs}, \emph{nectar}, \emph{ambrosia},
  \emph{fairies}'~\citep{maccoll1905}) constructed so as
  to `complete' the set of plausibility-indexed
  circumstances, \ie, in such a way that for each $\zqqq
  \equiv (\zq^k_i) \in \zQQQ$ (note the bar!)\
  there is always an $\zS_{\zqqq}$ --- possibly defined in
  terms of the fictitious $\set{\zSp_{j'}}$ --- for which
  $\pr(\zR_i \cond \zM^k \land \zSqq \land \zI) =
  \zq^k_i$. This operation --- which is, mark, not
  necessary --- has no importance nor mathematical
  consequences because the fictitious circumstances are
  impossible, \ie, their plausibilities in the context
  $\zI$ are naught, and thus terms containing them give no
  contribution in formulae like~\eqref{eq:decomp_general}
  or~\eqref{eq:Ri_cond_others}.}

Note that to obtain the marginal form above it is assumed
that knowledge of the measurement performed (but not of
its outcome!)\ is irrelevant for assigning the
plausibilities to the circumstances (\cf\
\eqn~\eqref{eq:ass_1_Ri}):
\begin{equation}
  \label{eq:M_redund}
  \pr(\zS_{\zqqq} \cond \zM^{k_1} \land \dotsb \land \zM^{k_n}\land \zI)
  =
  \pr(\zS_{\zqqq} \cond \zI)
\qquad
\text{for all $\zqqq$, $n=1,\dotsc\zm$, and $\set{k_t}$}.
\end{equation}

\section{Updating the plausibilities of circumstances
  and outcomes}
\label{sec:updating_gen}

Upon knowledge of the outcomes $\set{\zR^{k_1}_{i_1},
  \dotsc, \zR^{k_\zn}_{i_\zn}}$ of any subset
$\set{\zM^{k_1}, \dotsc, \zM^{k_\zn}}$, $\zn \le \zm$, of
measurements, the $\set{k_t}$ being all mutually different,
the plausibilities of the circumstances are updated,
with the assumption~\eqref{eq:M_redund}, according to
\begin{align}
  \label{eq:update_Sq_general}
  \pr[\zSqq \cond
(\zR^{k_1}_{i_1} \land\zM^{k_1}) \land \dotsb \land
  (\zR^{k_\zn}_{i_\zn} \land \zM^{k_\zn}) \land \zI]
&=
\frac{
 \zq^{k_1}_{i_1}\dotsm  \zq^{k_\zn}_{i_\zn} \, 
\pr(\zSqq \cond \zI)}
{\sum_{\zqqq' \in \zQQ}
 {\zq'}^{k_1}_{i_1}\dotsm  {\zq'}^{k_\zn}_{i_\zn} \, 
 \pr(\zSqql  \cond \zI)},
\intertext{or, in terms of the density $\zp$,}
  \label{eq:update_Sq_dens_general}
  \zp[\zqqq \cond
(\zR^{k_1}_{i_1} \land\zM^{k_1}) \land \dotsb \land
  (\zR^{k_\zn}_{i_\zn} \land \zM^{k_\zn}) \land \zI]
&=
\frac{
 \zq^{k_1}_{i_1}\dotsm  \zq^{k_\zn}_{i_\zn} \, 
\zp(\zqqq \cond \zI)}
{\int_{\zQQQ}
 {\zq'}^{k_1}_{i_1}\dotsm  {\zq'}^{k_\zn}_{i_\zn} \, 
 \zp(\zqqq'  \cond \zI)\,
\di\zqqq'}.
\end{align}

These formulae are valid for $\zn \ge 2$ only if we assume
that, when a circumstance is known and we want to assign a
plausibility distribution for a measurement, knowledge of
performance of other measurements or of their outcomes is
irrelevant (this is what Caves calls, in a slightly
different context (see the second paper in this series),
`learning through the parameter'~\citep{caves2000c}):
\begin{multline}
  \label{eq:ass_general_Ri}
  \pr(\zR^k_i \cond \zE \land \zM^k \land \zSqq \land\zI)
  = \pr(\zR^k_i \cond
  \zM^k \land \zSqq  \land\zI)
\\
  \mathbox{for all $\zqqq$, where $\zE$ is any conjunction
    of any number of mutually different $\set{\zM^{k_t}}$
    and any number of $\set{\zR^{k_t}_{i_t}}$ (each $k_t
      \neq k$).}
\end{multline}

Under the assumptions~\eqref{eq:M_redund}
and~\eqref{eq:ass_general_Ri}, we also obtain, by
marginalisation over the $\set{\zSqq}$, the plausibility
of an outcome $\zR^k_i$ given knowledge of outcomes of
other measurements $\set{\zM^{k_t}}$ different from
$\zM^k$:
\begin{equation}
  \label{eq:Ri_cond_others}
  \pr[\zR^k_i \cond
 (\zR^{k_1}_{i_1} \land\zM^{k_1}) \land
  \dotsb \land (\zR^{k_\zn}_{i_\zn} \land \zM^{k_\zn})
  \land \zM^k  \land\zI] 
=
\frac{
\int_{\zQQQ} \zq^k_i\,
 \zq^{k_1}_{i_1}\dotsm  \zq^{k_\zn}_{i_\zn} \,
\zp(\zqqq \cond \zI)\,\di\zqqq
}
{\int_{\zQQQ}
 {\zq'}^{k_1}_{i_1}\dotsm  {\zq'}^{k_\zn}_{i_\zn} \,  
 \zp(\zqqq'  \cond \zI)\,
\di\zqqq'}.
\end{equation}

\begin{rem}\label{rem:assumptions_notalways_hold}We should
  always be careful in assuming and using the conditions
  summarised in \eqns~\eqref{eq:M_redund}
  and~\eqref{eq:ass_general_Ri}, because they in many
  cases do not hold.
An example would be provided by the example of the coin
toss if we considered other tosses made by the
same, unknown, person. In the circumstance in which
\persc\ tosses the coin, \eqn~\eqref{eq:ass_general_Ri}
would not hold because from the results of other tosses we
would learn more about \persc's skills in coin-tossing. In
fact, even \eqn~\eqref{eq:persc} could cease to be valid
for other tosses, and our set of circumstances would no
longer be appropriate. We discuss similar matters in more
detail in the second part of this study.
In general, also the relations amongst the times or places
at which measurements are performed can be relevant and
thus require a careful analysis. \Cf\ the examples in
refs.~\citep{kirkpatrick2002b,kirkpatrick2001,kirkpatrick2002,mana2004,spekkens2004}.
\end{rem}

\section{Further remarks}
\label{sec:furth_remarks}

\begin{rem}
  A very important point is that the analysis of the
  context in terms of circumstances is far from unique
  (\cf\ footnote~\ref{fn:diff_circ}). Different sets
  $\set{\zSp'_{j'}}$, $\set{\zSp''_{j''}}$,
  $\set{\zSp'''_{j'''}}$, \etc\ of circumstances can be
  introduced to analyse the context, and from them
  corresponding sets of plausibility-indexed
  circumstances
  $\set{\zS'_{\zqqq} \st \zqqq \in \zQQ'}$,
  $\set{\zS''_{\zqqq} \st \zqqq \in \zQQ''}$,
  $\set{\zS'''_{\zqqq} \st \zqqq \in \zQQ'''}$, \etc\ can
  be constructed in the standard way.  The circumstances
  of each set have to be mutually exclusive and exhaustive
  for the present formalism to hold, but they need not be
  exclusive with those of the other sets. For example, in
  the case of the coin toss (\sect~\ref{sec:introexample})
  we could analyse the context $\zIc$ into another set of
  circumstances, say $\set{\zSp'_r \st r \in
    \opop{-1,1}}$ with
  \begin{equation}
    \label{eq:circ_weight}
    \zSp'_r \defd 
    \mathbox[.75]{\prop{The mass-centre of the coin lies on
        the coin's (oriented) axis a fraction $r/2$ of the total width away from the coin centre}.}
  \end{equation}
  The analysis and the construction of the plausibility-indexed
  circumstances would proceed exactly in the same way, apart from possibly
  different values of their plausibilities.\footnote{Note that the position
    of the mass-centre of a coin is not a very important factor in the
    assignment of a plausibility to heads or tails. See Jaynes'
    discussion~\citep[\chap~10]{jaynes1994_r2003}.}

  Different sets $\set{\zSp'_{j'}}$, $\set{\zSp''_{j''}}$,
  \ldots\ can also be combined into a single set with
  circumstances $\set{\zSp_{j'j''\dotso} \defd\zSp'_{j'}
    \land \zSp''_{j''} \land \dotsb}$. These will be
  mutually exclusive and exhaustive by
  construction. Again, the corresponding
  plausibility-indexed set $\set{\zS_{\zqqq}}$ will ensue
  in the usual way.
\end{rem}

\begin{rem}\label{rem:diff_priors_diff_circ}
  In view of the preceding remark it is clear that we can
  find \emph{a} meaning for a plausibility-parameter like
  $\zqqq$ in terms of a set of circumstances, but not
  \emph{the} meaning, because that set is not unique. This
  also implies that different choices of priors for
  $\zqqq$ need not be contradictory, because they can
  arise as the plausibilities for different sets of
  circumstances. There are, however, some compatibility
  conditions that the plausibility distributions for two
  or more sets of plausibility-indexed circumstances
  must satisfy (here stated in terms of densities):
\begin{multline}
    \label{eq:comp_cond_circ}
    \begin{aligned}
      \int\limits_{\zQQQ'} \zq^{k_1}_{i_1}\, \zpp(\zqqq \cond \zI)\,
      \di\zqqq &= \int\limits_{\zQQQ''} \zq^{k_1}_{i_1}\, \zppp(\zqqq
      \cond \zI)\, \di\zqqq,
\\
      \int\limits_{\zQQQ'} \zq^{k_1}_{i_1}\,  \zq^{k_2}_{i_2}\, \zpp(\zqqq \cond \zI)\,
      \di\zqqq &= \int\limits_{\zQQQ''} \zq^{k_1}_{i_1}\,
      \zq^{k_2}_{i_2}\, \zppp(\zqqq
      \cond \zI)\, \di\zqqq,
\\
      &\dots
\\
      \int\limits_{\zQQQ'} \zq^{1}_{i_1}
\,  \zq^{2}_{i_2}
\dotsm \zq^{\zm}_{i_\zm}\, \zpp(\zqqq \cond \zI)\,
      \di\zqqq &= \int\limits_{\zQQQ''} \zq^{1}_{i_1}
\,      \zq^{2}_{i_2} 
\dotsm \zq^{\zm}_{i_\zm}\,  \zppp(\zqqq
      \cond \zI)\, \di\zqqq,
    \end{aligned}
\\
\text{for all mutually different $k_t$ and
  appropriate $i_t$}
\end{multline}
(\ie, \emph{some} of their moments must be equal), where
$\zm$ is the number of measurements.  These conditions
arise simply analysing the plausibilities $\pr(\zR_i \cond
\zM^k \land\zI)$ and \eqref{eq:Ri_cond_others} first by
means of one set of circumstances, then by means of the
other, equating the expressions thus obtained, and
applying property~\eqref{eq:com_circ_lead_q} (under the
assumptions~\eqref{eq:M_redund}
and~\eqref{eq:ass_general_Ri}).
\end{rem}

\begin{rem}\label{rem:iteration_circ}
  The formalism lends naturally itself also to
  iteration. One can introduce `circumstances of
  circumstances', \etc, \ie\ deeper and deeper levels of
  analysis for the context $\zI$. What mathematically
  comes about looks like a hierarchy of `plausibilities
  of plausibilities', `plausibilities of plausibilities
  of plausibilities', \etc, which Good calls
  `probabilities of Type I, II, III',
  \etc~\citep{good1965}. Of course, such a cornucopia of
  recursive analyses may be appropriate and useful in some
  cases, while in others may just lead to constipation.
\end{rem}

\begin{rem}\label{rem:interval-valued_plaus}
  The interpretation here presented may also provide another point of view
  on theories of interval-valued probabilities (see
  \eg~\citep{atkinsonetal1964,kyburg1987}\citep[esp.\
  \sect~3.1]{ladetal1990} and
  \cf~\citep[\sect~2.2]{good1965}\citep{levi1984}), an in this sense
  completes or re-interprets studies by \eg\ Jamison~\citep{jamison1970},
  Levi~\citep{levi1974,levi1984}, Fishburn~\citep{fishburn1983},
  Nau~\citep{nau1992}.
\end{rem}

\chapter{Conclusions}
\label{sec:rem_gener}

We often have the need to use statistical models with
plausibility-like parameters, especially in classical and
quantum mechanics, and must face the problems of choosing
an suitable parameter space and a plausibility
distribution on this space. These problems would sometimes
be less difficult if the parameters could be given some
interpretation.

Some interpret the parameters as `propensities' or
`physical probabilities'. But these concepts do not make
sense to us. 

De~Finettians say that we should not interpret the
parameters, but think in terms of infinitely exchangeable
sequences instead; the parameters and their priors then
arise as mathematical devices. But we do not like being
forced to think in terms of infinite sequences, whose vast
majority ($\infty$) of elements must then necessarily be
fictitious.  And there are situations that can be repeated
a finite number of times only.

In addition to this, looking at concrete applications of
statistical models it seems that behind the parameters we
often have `at the back of our minds' an idea of some
possible hypotheses --- `circumstances' --- that could
hold in the context under study, \eg\ a physical
measurement. These circumstances could help us in the
assignment of plausibilities. And they need not concern
`causes' or `propensities'; see
remarks~\ref{rem:circum_no_causes}
and~\ref{rem:circumst_rem_2}. At the same time, we are
sometimes not interested in the intrinsic details of such
circumstances, but only in the plausibilities that we
eventually assign on their grounds.

We have seen in this study that plausibility theory allows
us, starting from \emph{any} set $\set{\zSp_j}$ of
circumstances, to form another, `coarse-grained' set
$\set{\zSq}$ with the property that its circumstances lead
each one to a different plausibility distribution. The
circumstances of this set can then be uniquely indexed by
the plausibility distributions they lead us to
assign. This set, moreover, is invariant with respect to
changes in the plausibilities of the initial and the
coarse-grained sets of circumstances, $\set{\pr(\zSp_j
  \cond \zI)}$ and $\set{\pr(\zSqq \cond \zI)}$.

This suggests that
plausibility-like parameters like $\zqq$, when used as
arguments of plausibility formulae, can always be
interpreted to stand for some appropriately indexed
circumstances like $\zSq$.  With mathematical care, this
may even hold for parameters of continuous statistical
models. Parameter priors like $f(\zqq \cond\zI)$ can
consequently be interpreted as plausibilities of
circumstances $\pr(\zSq \cond \zI)$.

The study of how these priors are updated when repetitions
of `similar' measurements occur, and of particular
applications to classical and quantum mechanics, are
developed in the next two papers.

\begin{acknowledgements}
  PM thanks Louise for continuous and invaluable
  encouragement, and the staff of the KTH
  \langswedish{Biblioteket} for their ever prompt support.
  AM thanks Anders Karlsson for encouragement.
\end{acknowledgements}

\newcommand{\bibpreamble}{Note: \texttt{arxiv} eprints are
  located at \url{http://arxiv.org/}.
}

\setlength{\bibsep}{0pt}

\newcommand{\bibfont}{\small}

\bibliography{bibliography}\end{document}

%% file: preamblemem.tex
\newcommand*\savesymbol[1]{%
  \expandafter\let\csname orig#1\expandafter\endcsname\csname#1\endcsname
  \expandafter\let\csname #1\endcsname\relax
}

\newcommand*\restoresymbol[2]{%
  \expandafter\global\expandafter\let\csname#1#2\expandafter\endcsname%
    \csname#2\endcsname
  \expandafter\global\expandafter\let\csname#2\expandafter\endcsname%
    \csname orig#2\endcsname
}

\usepackage[T1]{fontenc} 
\usepackage{textcomp}
\usepackage{amsmath}
\usepackage{amssymb}
\usepackage{amsxtra}
\usepackage[
]{babel}\FrenchItemizeSpacingfalse 
\newcommand{\langfrench}{\foreignlanguage{french}}
\newcommand{\langgerman}{\foreignlanguage{german}}

\newcommand{\langswedish}{\foreignlanguage{swedish}}
\newcommand{\langlatin}{\foreignlanguage{latin}}

\newcommand{\french}[1]{\emph{\langfrench{#1}}}

\savesymbol{qedsymbol}

\usepackage{amsthm}

\restoresymbol{ams}{qedsymbol}
\newcommand{\QEE}
{\amsqedsymbol}
\newcommand{\qee}{\leavevmode\unskip\penalty9999 \hbox{}\nobreak\hfill
\quad\hbox{\QEE}}

\usepackage[dvipdfm]{graphicx}

\providecommand{\firstname}[1]{#1}
\providecommand{\surname}[1]{#1}

\providecommand{\usetxfonts}{}
\usetxfonts

\newcommand{\Alpha}{\mathrm{A}}
\let\varAlpha A
\newcommand{\Beta}{\mathrm{B}}
\let\varBeta B
\newcommand{\Epsilon}{\mathrm{E}}
\let\varEpsilon E
\newcommand{\Zeta}{\mathrm{Z}}
\let\varZeta Z
\newcommand{\Eta}{\mathrm{H}}
\let\varEta H
\newcommand{\Iota}{\mathrm{I}}
\let\varIota I
\newcommand{\Kappa}{\mathrm{K}}
\let\varKappa K
\newcommand{\Mu}{\mathrm{M}}
\let\varMu M
\newcommand{\Nu}{\mathrm{N}}
\let\varNu N
\newcommand{\Omicron}{\mathrm{O}}
\let\varOmicron O
\newcommand{\Rho}{\mathrm{P}}
\let\varRho P
\newcommand{\Tau}{\mathrm{T}}
\let\varTau T
\newcommand{\Chi}{\mathrm{X}}
\let\varChi X
\newcommand{\omicron}{\mathrm{o}}

\DeclareFontFamily{U}{egreek}{\skewchar\font'177}%
\DeclareFontShape{U}{egreek}{m}{n}{<-6>s*[1]eurm5<6-8>s*[1]eurm7 <8->s*[1]eurm10}{}%
\DeclareFontShape{U}{egreek}{m}{it}{<->s*[1]eurmo10}{}%
\DeclareFontShape{U}{egreek}{b}{n}{<-6>s*[1]eurb5 <6-8>s*[1]eurb7 <8->s*[1]eurb10}{}%
\DeclareFontShape{U}{egreek}{b}{it}{<->s*[1]eurbo10}{}%
\DeclareSymbolFont{egreeki}{U}{egreek}{m}{it}%
\SetSymbolFont{egreeki}{bold}{U}{egreek}{b}{it}
\DeclareMathSymbol{\epartial}{\mathalpha}{egreeki}{"40}

\DeclareMathSymbol{\ealpha}{\mathalpha}{egreeki}{"0B}
\DeclareMathSymbol{\ebeta}{\mathalpha}{egreeki}{"0C}
\DeclareMathSymbol{\egamma}{\mathalpha}{egreeki}{"0D}
\DeclareMathSymbol{\edelta}{\mathalpha}{egreeki}{"0E}
\DeclareMathSymbol{\eepsilon}{\mathalpha}{egreeki}{"0F}
\DeclareMathSymbol{\ezeta}{\mathalpha}{egreeki}{"10}
\DeclareMathSymbol{\eeta}{\mathalpha}{egreeki}{"11}
\DeclareMathSymbol{\etheta}{\mathalpha}{egreeki}{"12}
\DeclareMathSymbol{\eiota}{\mathalpha}{egreeki}{"13}
\DeclareMathSymbol{\ekappa}{\mathalpha}{egreeki}{"14}
\DeclareMathSymbol{\elambda}{\mathalpha}{egreeki}{"15}
\DeclareMathSymbol{\emu}{\mathalpha}{egreeki}{"16}
\DeclareMathSymbol{\enu}{\mathalpha}{egreeki}{"17}
\DeclareMathSymbol{\exi}{\mathalpha}{egreeki}{"18}
\DeclareMathSymbol{\eomicron}{\mathalpha}{egreeki}{"6F}
\DeclareMathSymbol{\epi}{\mathalpha}{egreeki}{"19}
\DeclareMathSymbol{\erho}{\mathalpha}{egreeki}{"1A}
\DeclareMathSymbol{\esigma}{\mathalpha}{egreeki}{"1B}
\DeclareMathSymbol{\etau}{\mathalpha}{egreeki}{"1C}
\DeclareMathSymbol{\eupsilon}{\mathalpha}{egreeki}{"1D}
\DeclareMathSymbol{\ephi}{\mathalpha}{egreeki}{"1E}
\DeclareMathSymbol{\echi}{\mathalpha}{egreeki}{"1F}
\DeclareMathSymbol{\epsi}{\mathalpha}{egreeki}{"20}
\DeclareMathSymbol{\eomega}{\mathalpha}{egreeki}{"21}
\DeclareMathSymbol{\evarepsilon}{\mathalpha}{egreeki}{"22}
\DeclareMathSymbol{\evartheta}{\mathalpha}{egreeki}{"23}
\DeclareMathSymbol{\evarpi}{\mathalpha}{egreeki}{"24}
\let\evarrho\erho 
\let\evarsigma\esigma
\let\evarkappa\ekappa
\DeclareMathSymbol{\evarphi}{\mathalpha}{egreeki}{"27}
\DeclareMathSymbol{\evarAlpha}{\mathalpha}{egreeki}{"41}
\DeclareMathSymbol{\evarBeta}{\mathalpha}{egreeki}{"42}
\DeclareMathSymbol{\evarGamma}{\mathalpha}{egreeki}{"00}
\DeclareMathSymbol{\evarDelta}{\mathalpha}{egreeki}{"01}
\DeclareMathSymbol{\evarEpsilon}{\mathalpha}{egreeki}{"45}
\DeclareMathSymbol{\evarZeta}{\mathalpha}{egreeki}{"5A}
\DeclareMathSymbol{\evarEta}{\mathalpha}{egreeki}{"48}
\DeclareMathSymbol{\evarTheta}{\mathalpha}{egreeki}{"02}
\DeclareMathSymbol{\evarIota}{\mathalpha}{egreeki}{"49}
\DeclareMathSymbol{\evarKappa}{\mathalpha}{egreeki}{"4B}
\DeclareMathSymbol{\evarLambda}{\mathalpha}{egreeki}{"03}
\DeclareMathSymbol{\evarMu}{\mathalpha}{egreeki}{"4D}
\DeclareMathSymbol{\evarNu}{\mathalpha}{egreeki}{"4E}
\DeclareMathSymbol{\evarXi}{\mathalpha}{egreeki}{"04}
\DeclareMathSymbol{\evarOmicron}{\mathalpha}{egreeki}{"4F}
\DeclareMathSymbol{\evarPi}{\mathalpha}{egreeki}{"05}
\DeclareMathSymbol{\evarRho}{\mathalpha}{egreeki}{"50}
\DeclareMathSymbol{\evarSigma}{\mathalpha}{egreeki}{"06}
\DeclareMathSymbol{\evarTau}{\mathalpha}{egreeki}{"54}
\DeclareMathSymbol{\evarUpsilon}{\mathalpha}{egreeki}{"07}
\DeclareMathSymbol{\evarPhi}{\mathalpha}{egreeki}{"08}
\DeclareMathSymbol{\evarChi}{\mathalpha}{egreeki}{"58}
\DeclareMathSymbol{\evarPsi}{\mathalpha}{egreeki}{"09}
\DeclareMathSymbol{\evarOmega}{\mathalpha}{egreeki}{"0A} 
\DeclareSymbolFont{egreekr}{U}{egreek}{m}{n}%
\SetSymbolFont{egreekr}{bold}{U}{egreek}{b}{n}
\DeclareMathSymbol{\epartialup}{\mathalpha}{egreekr}{"40}

\DeclareMathSymbol{\ealphaup}{\mathalpha}{egreekr}{"0B}
\DeclareMathSymbol{\ebetaup}{\mathalpha}{egreekr}{"0C}
\DeclareMathSymbol{\egammaup}{\mathalpha}{egreekr}{"0D}
\DeclareMathSymbol{\edeltaup}{\mathalpha}{egreekr}{"0E}
\DeclareMathSymbol{\eepsilonup}{\mathalpha}{egreekr}{"0F}
\DeclareMathSymbol{\ezetaup}{\mathalpha}{egreekr}{"10}
\DeclareMathSymbol{\eetaup}{\mathalpha}{egreekr}{"11}
\DeclareMathSymbol{\ethetaup}{\mathalpha}{egreekr}{"12}
\DeclareMathSymbol{\eiotaup}{\mathalpha}{egreekr}{"13}
\DeclareMathSymbol{\ekappaup}{\mathalpha}{egreekr}{"14}
\DeclareMathSymbol{\elambdaup}{\mathalpha}{egreekr}{"15}
\DeclareMathSymbol{\emuup}{\mathalpha}{egreekr}{"16}
\DeclareMathSymbol{\enuup}{\mathalpha}{egreekr}{"17}
\DeclareMathSymbol{\exiup}{\mathalpha}{egreekr}{"18}
\DeclareMathSymbol{\eomicronup}{\mathalpha}{egreekr}{"6F}
\DeclareMathSymbol{\epiup}{\mathalpha}{egreekr}{"19}
\DeclareMathSymbol{\erhoup}{\mathalpha}{egreekr}{"1A}
\DeclareMathSymbol{\esigmaup}{\mathalpha}{egreekr}{"1B}
\DeclareMathSymbol{\etauup}{\mathalpha}{egreekr}{"1C}
\DeclareMathSymbol{\eupsilonup}{\mathalpha}{egreekr}{"1D}
\DeclareMathSymbol{\ephiup}{\mathalpha}{egreekr}{"1E}
\DeclareMathSymbol{\echiup}{\mathalpha}{egreekr}{"1F}
\DeclareMathSymbol{\epsiup}{\mathalpha}{egreekr}{"20}
\DeclareMathSymbol{\eomegaup}{\mathalpha}{egreekr}{"21}
\DeclareMathSymbol{\evarepsilonup}{\mathalpha}{egreekr}{"22}
\DeclareMathSymbol{\evarthetaup}{\mathalpha}{egreekr}{"23}
\DeclareMathSymbol{\evarpiup}{\mathalpha}{egreekr}{"24}
\let\evarrhoup\erhoup 
\let\evarsigmaup\esigmaup
\let\evarkappaup\ekappaup
\DeclareMathSymbol{\evarphiup}{\mathalpha}{egreekr}{"27}
\DeclareMathSymbol{\eAlpha}{\mathalpha}{egreekr}{"41}
\DeclareMathSymbol{\eBeta}{\mathalpha}{egreekr}{"42}
\DeclareMathSymbol{\eGamma}{\mathalpha}{egreekr}{"00}
\DeclareMathSymbol{\eDelta}{\mathalpha}{egreekr}{"01}
\DeclareMathSymbol{\eEpsilon}{\mathalpha}{egreekr}{"45}
\DeclareMathSymbol{\eZeta}{\mathalpha}{egreekr}{"5A}
\DeclareMathSymbol{\eEta}{\mathalpha}{egreekr}{"48}
\DeclareMathSymbol{\eTheta}{\mathalpha}{egreekr}{"02}
\DeclareMathSymbol{\eIota}{\mathalpha}{egreekr}{"49}
\DeclareMathSymbol{\eKappa}{\mathalpha}{egreekr}{"4B}
\DeclareMathSymbol{\eLambda}{\mathalpha}{egreekr}{"03}
\DeclareMathSymbol{\eMu}{\mathalpha}{egreekr}{"4D}
\DeclareMathSymbol{\eNu}{\mathalpha}{egreekr}{"4E}
\DeclareMathSymbol{\eXi}{\mathalpha}{egreekr}{"04}
\DeclareMathSymbol{\eOmicron}{\mathalpha}{egreekr}{"4F}
\DeclareMathSymbol{\ePi}{\mathalpha}{egreekr}{"05}
\DeclareMathSymbol{\eRho}{\mathalpha}{egreekr}{"50}
\DeclareMathSymbol{\eSigma}{\mathalpha}{egreekr}{"06}
\DeclareMathSymbol{\eTau}{\mathalpha}{egreekr}{"54}
\DeclareMathSymbol{\eUpsilon}{\mathalpha}{egreekr}{"07}
\DeclareMathSymbol{\ePhi}{\mathalpha}{egreekr}{"08}
\DeclareMathSymbol{\eChi}{\mathalpha}{egreekr}{"58}
\DeclareMathSymbol{\ePsi}{\mathalpha}{egreekr}{"09}
\DeclareMathSymbol{\eOmega}{\mathalpha}{egreekr}{"0A}

\newcommand{\alleugreek}{%
\let\alpha\ealpha
\let\beta\ebeta
\let\gamma\egamma
\let\delta\edelta
\let\epsilon\eepsilon
\let\zeta\ezeta
\let\eta\eeta
\let\theta\etheta
\let\iota\eiota
\let\kappa\ekappa
\let\lambda\elambda
\let\mu\emu
\let\nu\enu
\let\xi\exi
\let\omicron\eomicron
\let\pi\epi
\let\rho\erho
\let\sigma\esigma
\let\tau\etau
\let\upsilon\eupsilon
\let\phi\ephi
\let\chi\echi
\let\psi\epsi
\let\omega\eomega
\let\varepsilon\evarepsilon
\let\vartheta\evartheta
\let\varpi\evarpi
\let\varrho\evarrho 
\let\varsigma\evarsigma
\let\varkappa\evarkappa
\let\varphi\evarphi
\let\varAlpha\evarAlpha
\let\varBeta\evarBeta
\let\varGamma\evarGamma
\let\varDelta\evarDelta
\let\varEpsilon\evarEpsilon
\let\varZeta\evarZeta
\let\varEta\evarEta
\let\varTheta\evarTheta
\let\varIota\evarIota
\let\varKappa\evarKappa
\let\varLambda\evarLambda
\let\varMu\evarMu
\let\varNu\evarNu
\let\varXi\evarXi
\let\varOmicron\evarOmicron
\let\varPi\evarPi
\let\varRho\evarRho
\let\varSigma\evarSigma
\let\varTau\evarTau
\let\varUpsilon\evarUpsilon
\let\varPhi\evarPhi
\let\varChi\evarChi
\let\varPsi\evarPsi
\let\varOmega\evarOmega 
\let\alphaup\ealphaup
\let\betaup\ebetaup
\let\gammaup\egammaup
\let\deltaup\edeltaup
\let\epsilonup\eepsilonup
\let\zetaup\ezetaup
\let\etaup\eetaup
\let\thetaup\ethetaup
\let\iotaup\eiotaup
\let\kappaup\ekappaup
\let\lambdaup\elambdaup
\let\muup\emuup
\let\nuup\enuup
\let\xiup\exiup
\let\omicronup\eomicronup
\let\piup\epiup
\let\rhoup\erhoup
\let\sigmaup\esigmaup
\let\tauup\etauup
\let\upsilonup\eupsilonup
\let\phiup\ephiup
\let\chiup\echiup
\let\psiup\epsiup
\let\omegaup\eomegaup
\let\varepsilonup\evarepsilonup
\let\varthetaup\evarthetaup
\let\varpiup\evarpiup
\let\varrhoup\evarrhoup 
\let\varsigmaup\evarsigmaup
\let\varkappaup\evarkappaup
\let\varphiup\evarphiup
\let\Alpha\eAlpha
\let\Beta\eBeta
\let\Gamma\eGamma
\let\Delta\eDelta
\let\Epsilon\eEpsilon
\let\Zeta\eZeta
\let\Eta\eEta
\let\Theta\eTheta
\let\Iota\eIota
\let\Kappa\eKappa
\let\Lambda\eLambda
\let\Mu\eMu
\let\Nu\eNu
\let\Xi\eXi
\let\Omicron\eOmicron
\let\Pi\ePi
\let\Rho\eRho
\let\Sigma\eSigma
\let\Tau\eTau
\let\Upsilon\eUpsilon
\let\Phi\ePhi
\let\Chi\eChi
\let\Psi\ePsi
\let\Omega\eOmega
}

\newcommand{\unambeugreek}{%
\let\omicron\eomicron
\let\varAlpha\evarAlpha
\let\varBeta\evarBeta
\let\varEpsilon\evarEpsilon
\let\varZeta\evarZeta
\let\varEta\evarEta
\let\varIota\evarIota
\let\varKappa\evarKappa
\let\varMu\evarMu
\let\varNu\evarNu
\let\varOmicron\evarOmicron
\let\varRho\evarRho
\let\varTau\evarTau
\let\varChi\evarChi
\let\omicronup\eomicronup
\let\Alpha\eAlpha
\let\Beta\eBeta
\let\Epsilon\eEpsilon
\let\Zeta\eZeta
\let\Eta\eEta
\let\Iota\eIota
\let\Kappa\eKappa
\let\Mu\eMu
\let\Nu\eNu
\let\Omicron\eOmicron
\let\Rho\eRho
\let\Tau\eTau
\let\Chi\eChi
}

\usepackage{bm}
\providecommand{\piup}{\epiup}
\providecommand{\deltaup}{\edeltaup}
\providecommand{\zetaup}{\ezetaup}
\providecommand{\varepsilonup}{\evarepsilonup}

\providecommand{\coloneqq}{\mathrel{\mathop:}=}
\providecommand{\eqqcolon}{=\mathrel{\mathop:}}
\providecommand{\varprod}{\prod}

\newcommand{\delt}{\deltaup}
\newcommand{\celsius}{\textcelsius}
\newcommand{\di}{\mathrm{d}}

\DeclareSymbolFont{AMSb}{U}{msb}{m}{n}
\DeclareMathSymbol{\CC}{\mathalpha}{AMSb}{"43}
\DeclareMathSymbol{\RR}{\mathalpha}{AMSb}{"52}
\DeclareMathSymbol{\NN}{\mathalpha}{AMSb}{"4E}
\DeclareMathSymbol{\ZZ}{\mathalpha}{AMSb}{"5A}
\DeclareMathSymbol{\QQ}{\mathalpha}{AMSb}{"51}

\newcommand{\defd}{\coloneqq}
\newcommand{\defs}{\eqqcolon}

\newcommand{\Lor}{\bigvee}

\newcommand{\cond}
{\mathpunct{|}}
\newcommand{\st}{\mid}

\newcommand{\equi}{\sim}
\renewcommand{\le}{\leqslant}
\renewcommand{\ge}{\geqslant}
\DeclareMathDelimiter{\lclose}{\mathopen}{operators}{"5B}{largesymbols}{"02}
\DeclareMathDelimiter{\rclose}{\mathclose}{operators}{"5D}{largesymbols}{"03}
\DeclareMathDelimiter{\lopen}{\mathopen}{operators}{"5D}{largesymbols}{"03}
\DeclareMathDelimiter{\ropen}{\mathclose}{operators}{"5B}{largesymbols}{"02}
\newcommand{\clcl}[1]{\lclose#1\rclose}

\newcommand{\opop}[1]{\lopen#1\ropen}

\newcommand{\expe}[1]{\langle#1\rangle}
\newcommand{\set}[1]{\{#1\}}
\DeclareMathOperator{\pr}{P}
\newcommand{\prop}[1]{{\textnormal{`#1'}}}
\newcommand{\uni}[1]{\textnormal{#1}}
\newcommand{\sect}{\S}
\newcommand{\sects}{\S\S}
\newcommand{\chap}{ch.}%
\newcommand{\eqn}{eq.}%
\newcommand{\eqns}{eqs.}%

\theoremstyle{remark}
\newtheorem{remark}{Remark}
\theoremstyle{definition}

\newcommand{\etc}{{etc.}}
\newcommand{\ie}{{i.e.}}

\newcommand{\eg}{{e.g.}}
\newcommand{\viz}{{viz.}}
\newcommand{\cf}{{cf.}}
\newcommand{\Cf}{{Cf.}}

\newcommand{\etal}{{et al.}}

\newcommand{\bd}{\hspace{0pt}}%

\newcommand{\mathbox}[2][.8]{\parbox[t]{#1\columnwidth}{#2}}
\providecommand{\href}[2]{#2}
\providecommand{\eprint}[2]{\texttt{\href{#1}{#2}}}

\newcommand{\arxiveprint}[1]{eprint
  \texttt{\href{http://arxiv.org/abs/#1}{arXiv:\bd#1}}}

\newcommand{\citein}[1]{\textnormal{\citet{#1}}}